\documentclass{article}

\usepackage{arxiv}

\usepackage[utf8]{inputenc} 
\usepackage[T1]{fontenc}    
\usepackage{xcolor}
\usepackage[citecolor=blue,colorlinks=true,linkcolor=blue!50!black,urlcolor=blue!50!black]{hyperref}
\usepackage{url}            
\usepackage{booktabs}       
\usepackage{amsfonts}       
\usepackage{nicefrac}       
\usepackage{microtype}      
\usepackage{graphicx}
\usepackage[authoryear,round]{natbib}
\usepackage{doi}

\usepackage{pdflscape}
\usepackage{amsmath}

\title{The rise of populism and the reconfiguration of the German political space}

\author{ 
Eckehard Olbrich and Sven Banisch
    \\
	Max Planck Institute for Mathematics in the Sciences\\Leipzig, Germany\\
	\texttt{olbrich@mis.mpg.de} \\
}

\renewcommand{\headeright}{}
\renewcommand{\undertitle}{}
\renewcommand{\shorttitle}{Reconfiguration of the political space}

\hypersetup{
pdftitle={The rise of populism and the reconfiguration of the German political space},
pdfsubject={physics.soc-ph, cs.SI},
pdfauthor={Eckehard Olbrich, Sven Banisch},
pdfkeywords={political space, populism, topic model (LDA), party competition, network analysis},
}

\begin{document}
\maketitle

\begin{abstract}
The paper explores the notion of a reconfiguration of political space in the context of the rise of populism and its effects on the political system. We focus on Germany and the appearance of the new right wing party "Alternative for Germany" (AfD). The idea of a political space is closely connected to the ubiquitous use of spatial metaphors in political talk. In particular the idea of a “distance” between “political positions” would suggest that political actors are situated in a metric space. Using the electoral manifestos from the Manifesto project database we investigate to which extent the spatial metaphors so common in political talk can be brought to mathematical rigor. 

Many scholars of politics discuss the rise of the new populism in Western Europe and the US with respect to a new political cleavage related to globalization, which is assumed to mainly affect the cultural dimension of the political space. As such, it might replace the older economic cleavage based on class divisions in defining the dominant dimension of political conflict. An explanation along these lines suggests a reconfiguration of the political space in the sense that (1) the main cleavage within the political space changes its direction from the economic axis towards the cultural axis, but (2) also the semantics of the cultural axis itself is changing towards globalization related topics.

In this paper, we empirically address this reconfiguration of the political space by comparing political spaces for Germany built using topic modeling with the spaces based on the content analysis of the Manifesto project and the corresponding categories of political goals. 
We find that both spaces have a similar structure and that the AfD appears on a new dimension. In order to characterize this new dimension we employ a novel technique, inter-issue consistency networks (IICN) that allow to analyze the evolution of the correlations between the political positions on different issues over several elections. We find that the new dimension introduced by the AfD can be related to the split off of a new "cultural right" issue bundle from the previously existing center-right bundle. 

\end{abstract}


\section{Introduction}

\subsection{Political Spaces}
\label{intro:politcal_spaces}

Discussions about politics are full of spatial metaphors: one talks about political positions, about the distances between opinions or positions moving in a certain direction. Thus, it is natural to consider spatial representations of political opinions: political spaces. If one applies this idea to party competition, in the simplest case both a party and a voter have a position in this political space. The voter votes for the party with the position closest to their position while the party chooses a position that maximizes its share of votes. A first formalization of this idea as an economic model of spatial competition goes back to \citet{Hotelling1929}. A first explicit formulation as a model for competing political actors was given by Anthony \citet{Downs1957economic} in his economic theory of democracy  \cite[see][for a historical review]{Kurella2017evolution}. The model was originally formulated in a one-dimensional space, but later extended to more dimensions \citep{Stokes1963spatial}. \cite{Downs1957economic} explicitly thought about this space as an ideological space along the left-right axis. He motivated the use of such a low-dimensional space by the argument that the voters do not have information about the position of the parties or candidates on all the different political issues that are debated, because getting this information would be too costly. Instead, voters use the ideological position of the parties or candidates in a low dimensional space as a proxy for these positions. This is reasonable because the ideological position and the position on single issues are usually correlated. In this argument, Downs assumed the ideologies and the corresponding axis to be given. However, if these axes are not evident or even the matter of debate itself, one can turn the argument into empirical questions: How are the attitudes towards different political issues correlated for the actors (on the supply side) and which complexity reductions are actually used by the voters (on the demand side) to estimate the party preferences and to finally make their voting decisions. One can imagine that the agenda setting of mass media, but also attention cycles in the social media might have an influence on the latter and to some extent also the former. In this paper we will address these empirical questions only partially, i.e. on the supply side, by studying electoral manifestos of parties. We will do so by considering the inference of political spaces as task of dimension reduction \footnote{Our conceptualization is related to the idea of conceptual spaces developed by \cite{Gardenfors2000}. The link between political spaces and conceptual spaces in the sense of G\"ardernfors was already made in \cite{Laver2014}, but it remained abstract and was neither further elaborated nor used. Here we employ G\"ardenfors framework by considering the dimensions of the political space as higher order ``quality dimensions'' of the conceptual space with the ideological labels being higher order properties.}. 

The main level of political discourse that can be observed directly is the level of political issues. We consider anything on which a political decision can be made as a political issue. We can then consider a high-dimensional \emph{issue space} that is spanned by the possible attitudes on these issues.

Let us consider two concrete examples: Leuthold and Hermann \citep{Hermann2001,Leuthold2007} analyzed the votes of the Swiss population between 1980 and 2000. In this period people had to vote on 158 different political proposals. Leuthold and Hermann assigned these proposals to political goals on three levels of abstraction: tangible goals (20), institutional goals (9) and value oriented goals (14). Examples are ``animal protection'', ``direct democracy'' or ``law and order'', respectively. It is clear that these goals are categories that comprise each different political issues. In this work the list of political goals was determined by the issues that were addressed in the proposals in this period.  

As a second example let us consider the Manifesto Research on Political Representation (MARPOR) \citep{Manifesto2020-2, Merz2016}. In this project electoral manifestos of political parties from all over the world are collected and their content is encoded by trained native-language experts. They split the programs into statements (so-called quasi-sentences) and assign to each statement a category from a list of policy goals. Again these categories have a multi-level structure, with the main level comprising 56 categories, which are aggregated into 7 fields of politics (freedom and democracy, social policy, economic policy, government organization, foreign affairs, cultural policy, appeal to target groups).

For a single actor, such as a politician, a political party, or a voter, the attitude towards these issues is represented as a point (or region) in this issue space. If, in a next step, a population of such actors is considered, the attitudes on the single issues are usually correlated due to logical, psychological and social constraints \citep{Converse1964nature}, which allows to represent the political differences between these actors in low dimensional latent spaces that are called ``political spaces'' \citep{Laver2014}. Traditionally their dominant dimension is denoted by left versus right. But, in contrast to the issue space, the political space is a \emph{latent space} and its dimensions cannot be observed directly, which is reflected in the often confusing discussion about the meaning of the ideological labels ``left'' and ``right''. 
In this framework the meaning of the dimensions of the political space has to be inferred empirically, can change over time and is only given with respect to the considered population. This approach is sometimes also called the inductive approach to political spaces \citep{Daeubler2017estimating}. As already noted, inductively derived political spaces are relative to the population under consideration. Whether different populations (e.g. regional vs. national vs. European) give rise to similar political spaces will then be an empirical question.

In contrast to the inductive approach one can also following a deductive approach by deriving a lower dimensional space analytically from certain  basic distinctions such as equality-inequality in the case of the left-right distinction \citep{Bobbio1996left}. Attitudes towards specific political issues are then considered as consequences of attitudes towards a set of basic values or distinctions\footnote{Mathematically both approaches define projections from the issue space towards a lower dimensional subspace.}. There is an ongoing debate between a deductive \citep{Budge2013standard} and an inductive approach \citep{Gabel2000putting,Daeubler2017estimating} to construct political spaces. The proponents of an inductive approach argue that the meaning of political labels such as ``left'' and ``right'' or ``liberal'' and ``conservative'' is not constant over time. Scholars following a deductive approach on the other side argue that exactly such changes of ideological positions over time can be seen only with respect to a constant frame of reference. 

In this paper, we study how the emergence of new parties are reflected in the political space. As we will elaborate in the next section, the different hypotheses may relate to both changing positions in an existing political space, but also to a change in the composition of the axes itself. Thus we will use both deductive and inductive political spaces and combine them with a new technique to construct ``issue bundles'' (see sec. \ref{method:issue_bundles}). We think that inductive techniques are inevitable, in order to explore new phenomena, otherwise one could only confirm existing expectations. 
The distinction between deductive and inductive approaches can be made not only for the construction of the political space by dimension reduction, but already for defining the political issues. If one measures attitudes on political issues, be it with surveys or by content analysis as in the Manifesto project, one usually works with predefined categories of issues. However, also these categories could undergo semantic changes over time or might need refinement. An example would be, that the manifesto project refined there categories in version 5 of their codebook \citep{Manifesto2020-codebook}, for refining categories that are in particular relevant for the characterization of populist parties, such as democracy, distinguishing now between representative and direct democracy, or introducing immigration as a subcategory to multiculturalism. In this paper we will explore topic modelling as a fully inductive, unsupervised form of content analysis for generating the issue space.

\subsection{The rise of populist movements and parties and the new cleavage}
\label{intro:new_cleavage}

A popular explanation for the latest wave of electoral successes of populist candidates, movements or parties is the idea of a representation gap \citep{Hall2019}. 
It basically means that the parties have changed their political positions in a way that certain parts of the electorate are not represented anymore, i.e. the parties have moved away from them. This narrative comes in two versions: one for center-right (e.g. Christian Democratic) parties and one for the center-left, i.e. the social-democratic parties.  
\begin{enumerate}
        \item The center-right parties moved to the left and created a ``gap'' on the conservative side of the political spectrum that was filled by new right parties. This refers in Germany mostly to non-economic issues such as nuclear phase out (Atomausstieg), suspension of compulsory military service or gay marriage.
        \item For the center-left parties there are again two versions. In both versions the diagnosis is that they neglected the fight for the economic interests of the less privileged. The first narrative is older and criticizes that in particular the social democrats have moved to the right with respect to economic issues by adopting ``neoliberal'' politics. Examples would be the politics of ``New Labour'' under Tony Blair in the UK and of the German social democrats under Gerhard Schröder and their orientiation towards a ``new center''  \citep{blair1999europe}. After the Brexit referendum and Trump's election 2016 a second narrative became more prominent (see for instance \citet{lilla2018once}): that the left has changed their focus towards cultural issues and the so-called ``identity politics'' which had led to a loss of the ``working class'' voters.
\end{enumerate}

These narratives are mainly told by political commentators, pundits or politicians themselves. In the academic literature on comparative politics scholars try to understand these phenomena in the framework of cleavage theory \citep{LipsetRokkan1967}, which states that the main lines of political conflicts are  related to a few cleavages. A full cleavage consists of three elements \citep{bartolini1990policy}: (1.) a difference in social structure, (2.) common interests or values creating a sense of collective identity and (3.) political alternatives in the form of voting alternatives, i.e. parties. The three elements correspond to three layers of description: (1.) the socio-demographic structure (2.) the prevalent attitudes, narratives and ideologies and (3.) the political system and its institutions. While older theories mainly considered a bottom-up causality, for instance from the social structure to the attitudes to party positions, in more recent work the relative autonomy of the layers is more emphasized and there is also research on instances of top-down causation: Parties might not only accentuate or de-emphasize existing structural or attitudinal differences, but they might even produce new ones by forming communities or creating novel identities \citep{DeeganKrause2007}. \citet{LipsetRokkan1967} used cleavage theory to explain the formation of the party systems in western Europe starting from the industrial and national revolution of the 19th century. They considered 4 cleavages: (1.) center vs. periphery as conflicts about the form of the national state, (2.) state vs. church as conflict about the secularization of, for instance, the education system, (3.) the conflict between interest of the agrarian sector and the industrial sector, and finally (4.) the most formative conflict - the class conflict between the working class and the owning class. While these 4 cleavages would correspond to a political space with 4 dimensions, many scholars considered two-dimensional spaces with an economic axis corresponding to the class cleavage and a cultural axis representing conflicts related to one or more of the other cleavages \citep[see for example][]{Hix1999,Kriesi2006,Bakker2013}. This can work because those conflicts are usually not independently articulated, but often the same parties take side on different conflicts.  

When Lipset and Rokkan formulated their theory, they had the impression that the party systems in western democracies had been relatively stable: "the party systems of the 1960s reflect, with a few but significant exceptions, the cleavage structure of the 1920s.", i.e. the party system was ``frozen'' \citep{Schoultz2017party}.  But at the very time they formulated this hypothesis, the several new phenomena started to kick in. New social movements started to form, such as the civil rights movement, peace movement, or ecological movements. New parties started to emerge, first on the left with green and alternative left parties, later also "new right" and anti-immigration parties. And the support of the large center-left, the social democrats, and center-right parties started to decay. Within the framework of cleavage theory this was conceptualized as a ``new cleavage'' \citep{Kitschelt1990,Kriesi2006,Bornschier2010,Oesch2012} related to the transition from industrial societies to post-industrial societies that goes with structural changes that could be observed in all western democracies: the education expansion, the feminization of the job market, higher geographic mobility, or more migration. Moreover, also trends that started already during industrialization were persisting, such as urbanisation and secularization. On the attitudinal level these processes were linked to processes of value change, the "silent revolution" \citep{Inglehart1977} from values emphasizing material wealth and physical security towards  "postmaterial" values related to social, cultural, and intellectual needs. In more recent work, \cite{inglehart2005modernization} distinguish two processes of value change: one related to the transition from agrarian to industrialized societies, giving rise to secular-rational values, and another one linked to the emergence of post-industrial societies and the increasing prevalence of "self-expression values" that are strongly related to liberal attitudes on socio-cultural issues such as abortion, legalization of drugs or gay rights. On this basis, \citet{Norris2019} developed their "cultural backlash" thesis which explains the most recent rise of populist movements and parties as a reaction to the "postmaterial" values becoming hegemonic in western societies, and many attitudes that were considered to be "normal" in the past are now questioned and are coming under pressure to justify\footnote{That is also reflected in the slogan of the AfD for the national elections in this year: "Deutschland. Aber normal" - "Germany. But normal".}. \citet{Piketty2018} and \citet{Gethin2021} emphasize the importance of education for the "new cleavage" between the green and  new left parties on the one side and "anti-immigration" parties on the other side. They show that the former are overproportionally supported by the more educated while they found the opposite picture for the latter. Also \citet{Norris2019} identify the education expansion, i.e. the increasing share of the population with higher education, as one of the driving forces of the value change. 
While the traditional center-left and center-right -- the SPD and the CDU/CSU in Germany -- represented the two sides of the traditional class cleavage, the new cleavage led not only to the emergence of new parties, but it is also increasingly affected the voter bases of these traditional parties by creating diverging interests. Both developments sketched above - the Social Democrats trying to reach out to a ``new center'' and the Christian Democrats liberalizing some of their positions on socio-cultural issues - can be interpreted as an attempt to react to this new situation. 

Another line of reasoning focuses more on different interests with respect to globalization processes by contrasting those that will profit from globalization, or at least have the resources to cope with the associated challenges, and those that perceive or fear a loss of their familiar world. Popular labels for the two sides are "somewheres" versus "anywheres" \citep{goodhart2017road} or "cosmopolitans" versus "communitarians" \citep{deWilde2019struggle}. 
A comprehensive discussion of these theses and the underlying theories is beyond the scope of this paper, because we do not address here the changes on the side of the voters, at least not directly. Instead, we are interested to which extent we can observe a change in the programmatic structure of the party manifestos which would be reflected in a reconfiguration of the political space as it is predicted by the ``new cleavage'' in the sense of a "re-bundling" the issues. Moreover, there is also the possibility that new competitors in the party system do not simply occupy empty spaces in the existing party system, i.e. the political space spanned by the differences between the existing parties. Instead, they might present also new combinations of political goals, new ``issue bundles'' \citep[p. 60]{Daeubler2017estimating}. In the framework of cleavage theory \cite{Hooghe2018} argue that party system change comes in the form of rising parties because of issue coherence and programmatic stickiness on the side of the existing parties. It would be interesting to see how that might add new dimensions to the political space on the one hand, and whether it is followed by an adaptation process both from the side of the challengers and the challenged that would eventually lead to an incorporation of these new dimensions into the existing political order. 

Our aim in this paper is to develop and improve the computational tools for the observation and analysis of political positions of parties based on their electoral manifestos and to use them to explore the case of the new right-wing populist party ``Alternative for Germany'' (AfD) in Germany. The paper consists of two main parts. In the first part (section \ref{sec:data_methods}) we present the data and methods used in this paper. In particular, we discuss  methods for estimating political spaces (section \ref{meth:political_spaces}) and for constructing ``issue bundles'' or meta-issues from data (section \ref{method:issue_bundles}) and show first examples for our data set. In the second part (sections \ref{sec:results} and \ref{sec:discussion}) we apply our methods to our case study: the development of the German political parties since 1990 and the appearance of the AfD. In particular we want to address the following questions:

\begin{enumerate}
	\item To which extent can we define a meaningful left-right distinction as the main difference in politics, how stable is it and has it changed its meaning?
	\item Can we find evidence for a representation gap in the political spaces preceding the appearance of the national populist parties? 
	\item Do we find evidence for specific changes of the political space in connection with the appearance of these parties? 
\end{enumerate}

\section{Data and Methods}
\label{sec:data_methods}
\subsection{Manifesto data}

The Manifesto Research on Political Representation (MARPOR) project \citep{Merz2016} collects electoral manifestos of political parties from all over the world and has their content encoded by trained native-language experts. In this paper we used the most recent version of the dataset \citep{Manifesto2020-2}. The coders split the programs into statements (so-called quasi-sentences) and assign to each statement a category of an extensive coding scheme of policy goals \citep{Manifesto2020-codebook}. We used the version 4 of the coding scheme that comprises 56 substantial categories and one category for unencoded quasisentences. The 56 categories are grouped into 7 domains: External relations (10 categories), Freedom and Democracy (4), Political System (5), Economy (16), Welfare and Quality of Life (7), Fabric of Society (8) and Social Groups (6). By counting the frequency of occurrences of certain political goals in one document, each electoral manifesto can be represented by the normalized counts. The normalization is done with respect to the total number of encoded quasi-sentences. Thus, the normalized count can be considered as an empirical probability. Many categories differentiate between a positive and a negative attitude towards the same political issue. Examples are ``Military:positive'' and ``Military:negative'' or ``European union:positive'' and ``European union:negative''. But, in agreement with saliency theory of party competition \cite{Budge1982}, which states that parties tend to emphasize the issues they own and are less likely seeking direct confrontation - at least in their electoral manifestos - very often, only one pole is significantly expressed in a single election. 

\subsection{Topic modeling}
\label{meth:topic_models}
For a large subset of the electoral manifestos the Manifesto project also provides the text of the manifestos in a machine readable form and for a smaller subset also the pairs of quasi-sentences and the assigned codes for the political goals. These data are available via API from the Manifesto project webpage. We trained two topic models:
\begin{enumerate}
	\item[\textbf{A}] One topic model with 60 topics (Tables \ref{tab:topic_model_A1},\ref{tab:topic_model_A2}) for all programs from Germany for the elections from 1949 - 2017.
	\item[\textbf{B}] One topic model with 40 topics (Table \ref{tab:topic_model_B}) for all annotated programs for election from 1998 - 2017.
\end{enumerate}

API calls and the pre-processing of the texts including the lemmatization was done in Python using SpaCy \citep{spacy} for the lemmatization and Gensim \citep{gensim} for the bigram detection. For the topic modeling we used the standard latent Dirichlet allocation (LDA) model \citep{Blei2003}. Before estimating the topic model the manifestos were split into parts of approximately 100 tokens each. This was done because using the whole program as a single document would provide only a very small number of documents, For instance, in the case of Germany, there are in total 89 programs for the elections from 1949 to 2017. Thus, one had to restrict oneself to either a small number of topics in order to avoid the trivial solution of one topic per document, or one has to use the prior to ensure broad topic distributions for all programs. The topic model was estimated using the Text Analysis toolbox from Matlab using the collapsed variational Bayes algorithm including fitting of the topic concentration. The optimal number of topics was determined by crossvalidation using 10 \% of the corpus as test set and 90 \% for training. By averaging the topic probabilities of all documents that belong to a single electoral manifesto we can again characterize each manifesto by probability vector, but in contrast to the MARPOR data set the probabilities now represent topics and not categories of political goals.     

The simplest way to select labels for a topic is to take the words with the largest probabilities. However, there might be high probability words that appear in many topics. Thus, topic labels should be also specific for the labeled topic. \citet{sievert2014ldavis} proposed to use for this purpose the relevance of a word for a topic which is defined as 
\begin{align}
	r(w,t|\lambda)&=\lambda log(\phi(w|t)) + (1-\lambda) \log \frac{\phi(w|t)}{p(w)}  \nonumber \\
	&=\log \phi(w|t) - (1-\lambda) \log p(w)
	\label{eq:relevance}
\end{align}
with $p(w)$ being the overall word probability and  $\phi(w|t)$ the word probabilities per topic. The first term is the log probability of the word in the topic while the second term is the pointwise mutual information between the topic and the word, which quantifies how informative the word is for identifying the topic. For creating the English topic labels we looked at the three words or bigrams with the highest relevance using $\lambda =\{0,0.3,1\}$. From these words we tried to find a label that best summarizes the content of the topic. In difficult cases, we looked also on more words. Tables \ref{tab:topic_model_A1},\ref{tab:topic_model_A2} and \ref{tab:topic_model_B} show for both topic models the most relevant words for $\lambda=1$ (most probable in topic) and $\lambda=0.3$ (selects more topic specifc words) and the English label. 

\subsection{Political spaces}
\label{meth:political_spaces}
\begin{figure}[h!]
\centering
	\includegraphics[width=0.8\linewidth]{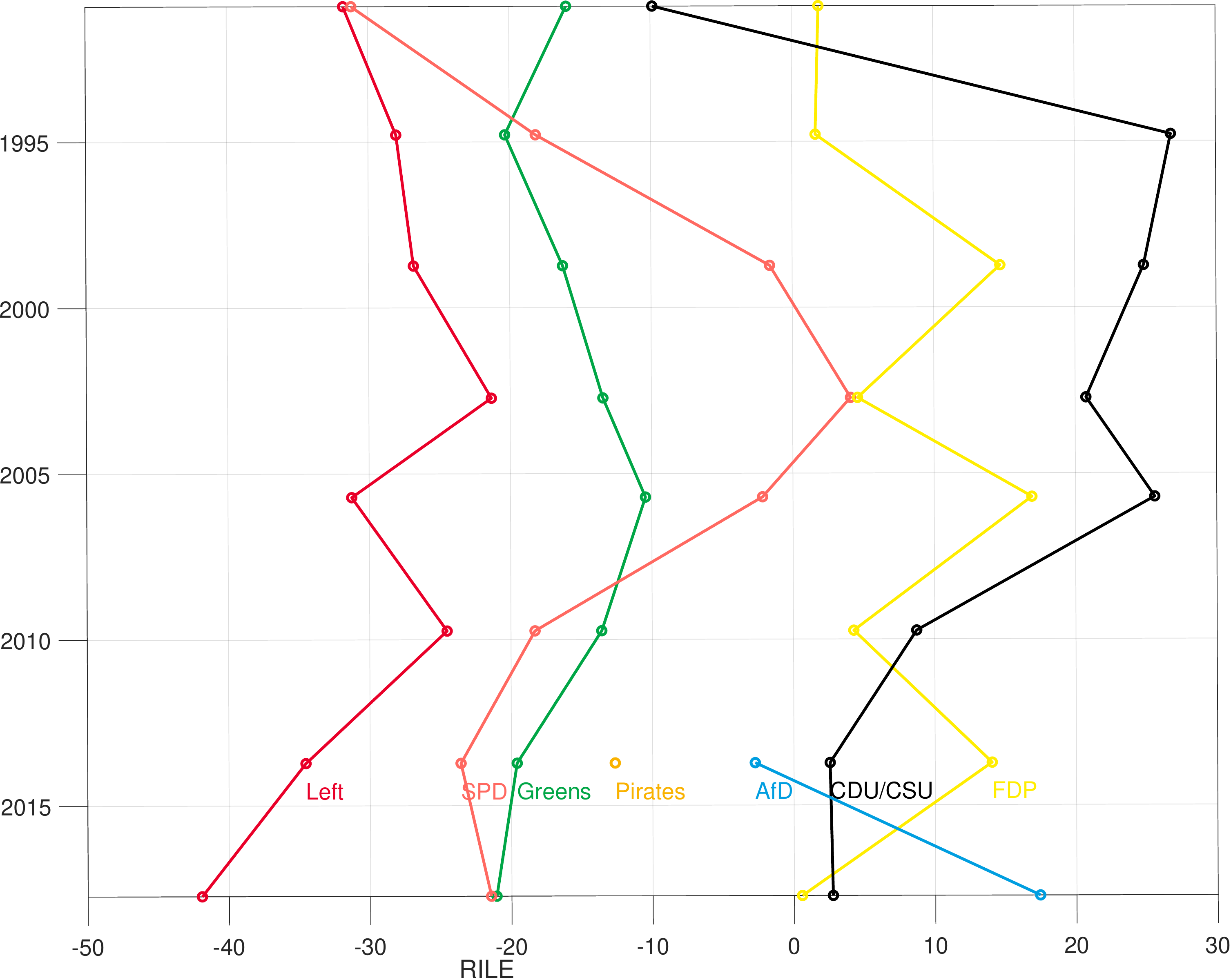}
	\caption{RILE index of the German parties between 1990 and 2017.}\label{fig:rile}
\end{figure}

The most widely used  one-dimensional political space that is used with MARPOR data is the RILE index, a left-right scale that is estimated by calculating the difference between the sum of the normalized counts (in percent) of a set of 13 left and right issues, respectively. Thus, it lies between +100 for a program containing only right and -100 for program with only left issues. Fig.~\ref{fig:rile} shows the RILE index for the German parties between 1998 and 2017. Table \ref{tab:RILE} shows which issues are considered as right and left, respectively. This index was constructed originally based on the programs of western Europan parties 1945-1985 by a multistage process starting from an exploratory factor analysis in single countries \citep{Laver1992measuring}, combined with theoretical consideration. Thus, it was originally at least partially constructed by an inductive approach, but is since then mostly considered as a deductive political space. 

\begin{table}[t]
	\centering
	\begin{tabular}{|l|l|} \hline
		Right & Left \\ \hline
		Military: Positive (104) & Anti-imperialism (103)\\
		Freedom and Human Rights (201) & Military: Negative (105)\\
		Constitutionalism: Positive (203) & Peace (106)\\
		Political Authority (305) & Internationalism: Positive (107)\\
		Free Market Economy (401)& Democracy (202)\\
		Economic Incentives (402) & Market Regulation (403)\\
		Protectionism: Negative  (407) & Economic Planning (404)\\
		Economic Orthodoxy (414) & Protectionism: Positive (406) \\
		Welfare State Limitation (505) & Controlled Economy (412)\\
		National Way of Life: Positive (601) & Nationalisation (413)\\
		Traditional Morality: Positive (603) & Welfare State Expansion (504)\\
		Law and Order (605) & Education Expansion (506)\\
		Civic Mindedness: Positive (606) & Labour Groups: Positive (701) \\ \hline
	\end{tabular}
	\caption{Issues used for estimating the RILE index \citep{Budge2013standard}.}
	\label{tab:RILE}
\end{table}

\begin{figure}[h!]
	\includegraphics[width=\linewidth]{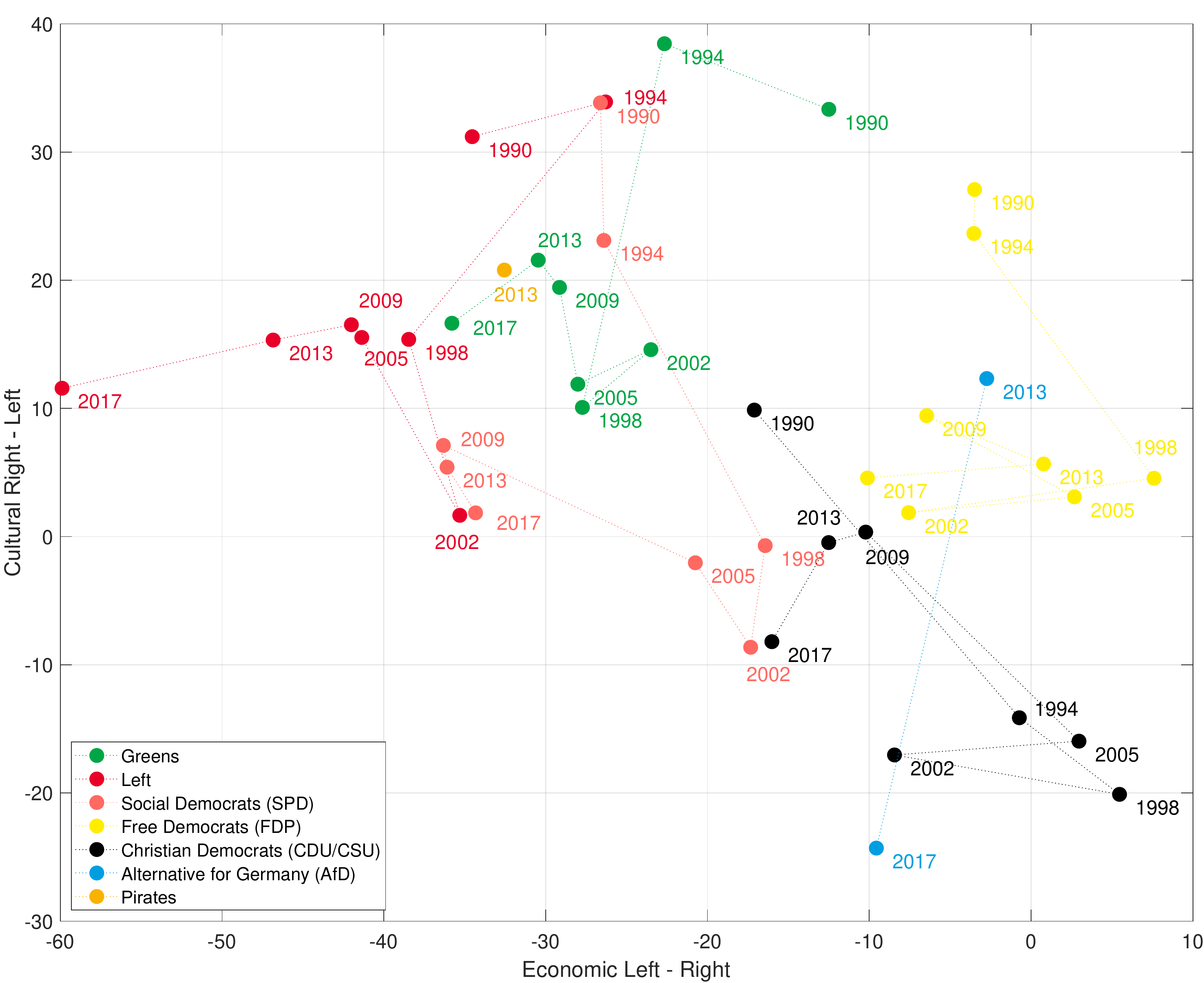}
	\caption{Projection on the 2-d political space with the economic and cultural axis spanned by the difference between the economic and cultural left and right categories, respectively (see Tab.~\ref{tab:BakkerHobolt}) of the German parties between 1998 and 2017.}\label{fig:BakkerHobolt}
\end{figure}

In order to capture also changes in the party system related to the ``new cleavage'' shortly sketched in sec.~\ref{intro:new_cleavage}, \citet{Bakker2013} proposed a two-dimensional space with an economic and a socio-cultural dimension. They called the poles of the socio-cultural axis "Libertarian" and "Authoriarian". See table~\ref{tab:BakkerHobolt} for the corresponding MARPOR categories. We use here less specific labels, following \cite{Gethin2021} to avoid too narrow associations. 

\begin{table}
	\centering
	\begin{tabular}{|l|l|} \hline
		Economic left & Economic right \\ \hline
		Market Regulation (403)   & Free Market Economy (401) \\
		Economic planning (404)		& Incentives: positive (402)  \\
		Corporatism: positive (405) & Protectionism: negative (407) \\
		Protectionism: positive (406)  & Economic growth positive (410) \\ 
		Keynesian demand management (409) & Economic orthodoxy (414) \\
		Controlled economy (412) & \\
		Nationalization (413) & \\  
		Equality (503) & \\ 
		Welfare state expansion (504) & Welfare state limitation (505) \\
		Educational expansion (506) & Educational limitation (507) \\
		Labour groups: positive (701) & Labour groups negative (702) \\ \hline
		Socio-cultural left & Socio-cultural right \\ \hline
		Democracy (202) & Political authority (305) \\
		Environmental protection (501) & \\
		Culture: positive (502)& \\
		National way of life: negative (604)& National way of life: positive (601)\\
		Traditional Morality: negative (604) & Traditional morality: positive (603) \\
		& Law and order: positive (603) \\
		& Civic mindedness: positive (606)\\
		Multiculturalism: positive (607) & Multiculturalism: negative (608) \\
		Underprivileged minority groups (705) & \\
		Non-economic demographic groups (706) & \\ \hline
	\end{tabular}
\caption{Issues used to define an economic and a socio-cultural dimension in \cite{Bakker2013}.}
\label{tab:BakkerHobolt}
\end{table}

\begin{figure}[h!]
	\includegraphics[width=0.48\linewidth]{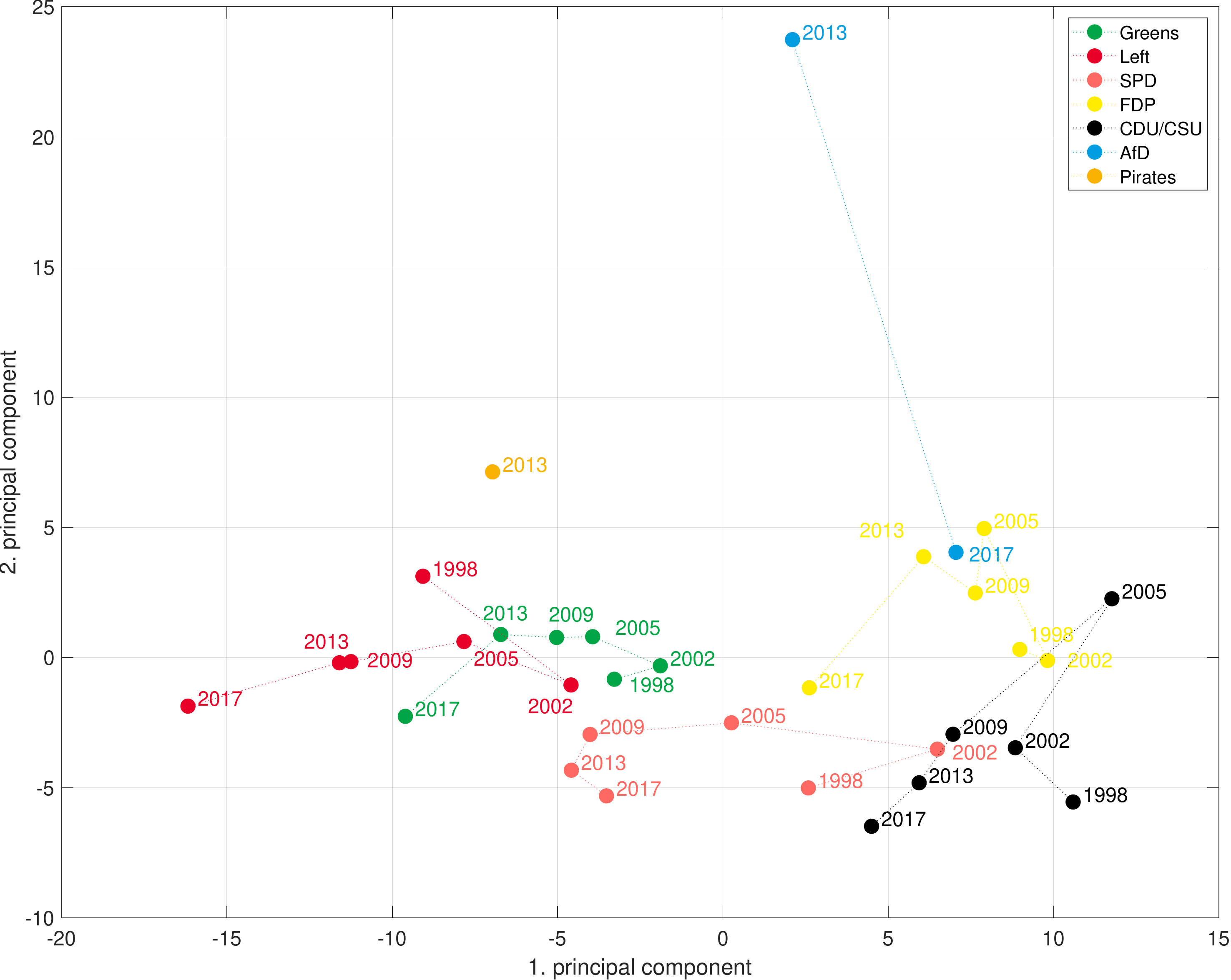}
	\includegraphics[width=0.48\linewidth]{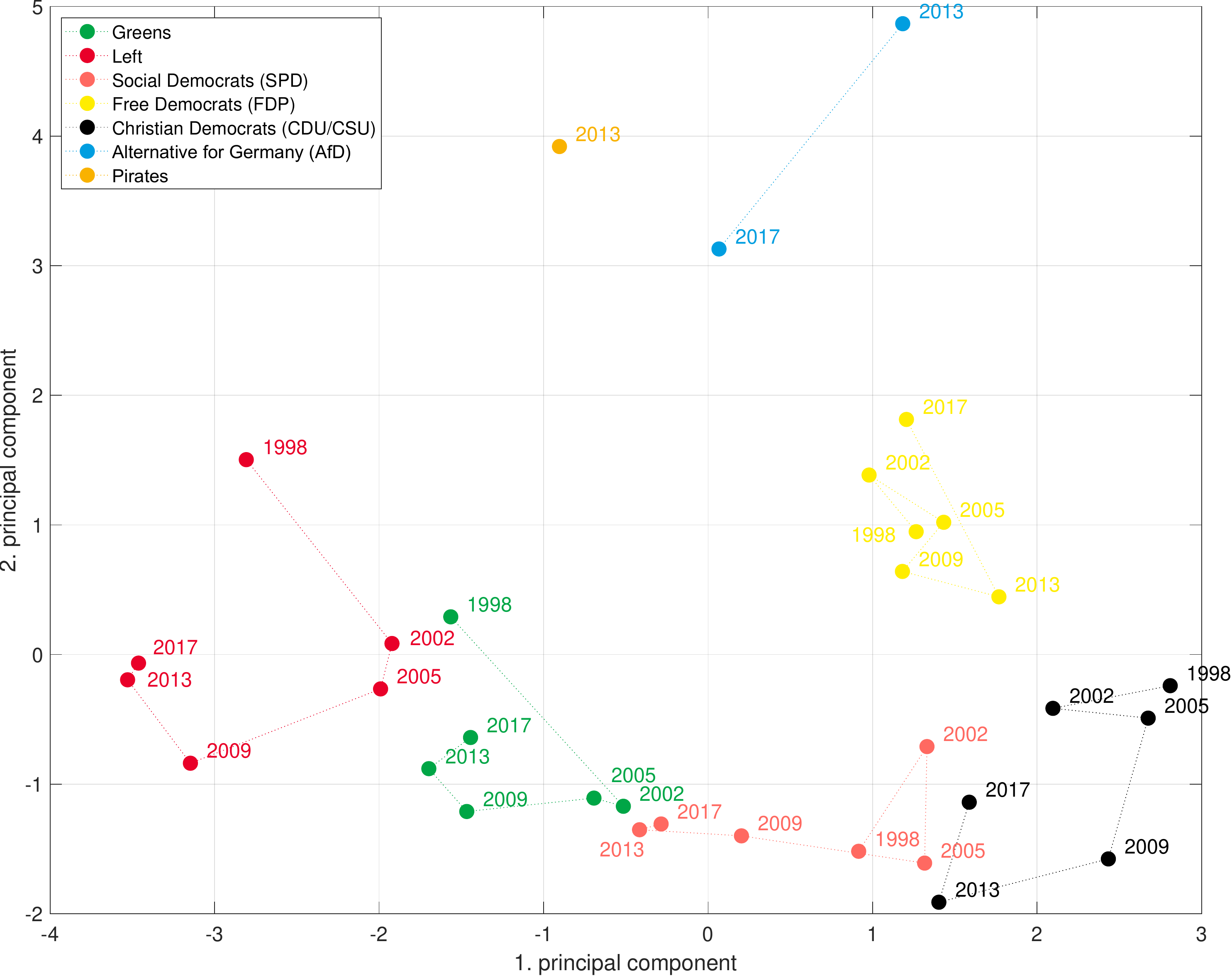}
	\caption{Political spaces spanned by the first and second PCA component from the MARPOR counts (left) and for the topic model \textbf{B} (right) for Germany 1998 - 2017.}\label{fig:pca_1998_2017_1_2}
\end{figure}

Fig.~\ref{fig:BakkerHobolt} shows the space for Germany 1998 - 2017. We will use these two deductive spaces, one dimensional RILE space and the two dimensional economic-cultural space, as references for our inductively constructed spaces. For constructing the latter we use a principal component analysis with utilizing the vectors of issue probabilities in the case of the MARPOR data and the vector of topic probabilities in the case of the topic model. We also considered to weight the manifestos with the vote share at the corresponding elections to avoid that the spaces are dominated by small parties, but decided against it, because first, we want to consider political spaces as semantic spaces and, second, for the cases that we study in this paper, some of the political proposals of smaller parties, such as the AfD (Alternative for Germany) or the Pirates appeared to be much more salient in the public discussion than it was reflected in there final vote share (4.7\% and 2.2\%, respectively).
The scores provide the coordinates of the party manifestos in the space spanned by the principal components, while the loadings allow to interpret the axis in terms of the categories or topics. 
Such political spaces can be estimated for single elections or for longer periods. Figure~\ref{fig:pca_1998_2017_1_2} shows the inductive political spaces spanned by the two first principal components for the German electoral manifestos between 1998 and 2017. 

\subsection{Issue bundles}
\label{method:issue_bundles}

A better understanding of the structure of the ideological space in which the political competition takes place can be obtained by analysing directly systematic differences across the entire set of parties that compete for votes in a given election. If, for instance, a subset of parties scores high on a certain selection $A$ of categories but low on another selection $B$, whereas other parties scores low on the former and high on the latter, this points towards bundles of categories that are either jointly supported or dismissed.

\begin{figure}[ht]
	\centering
	\includegraphics[width=0.99\linewidth]{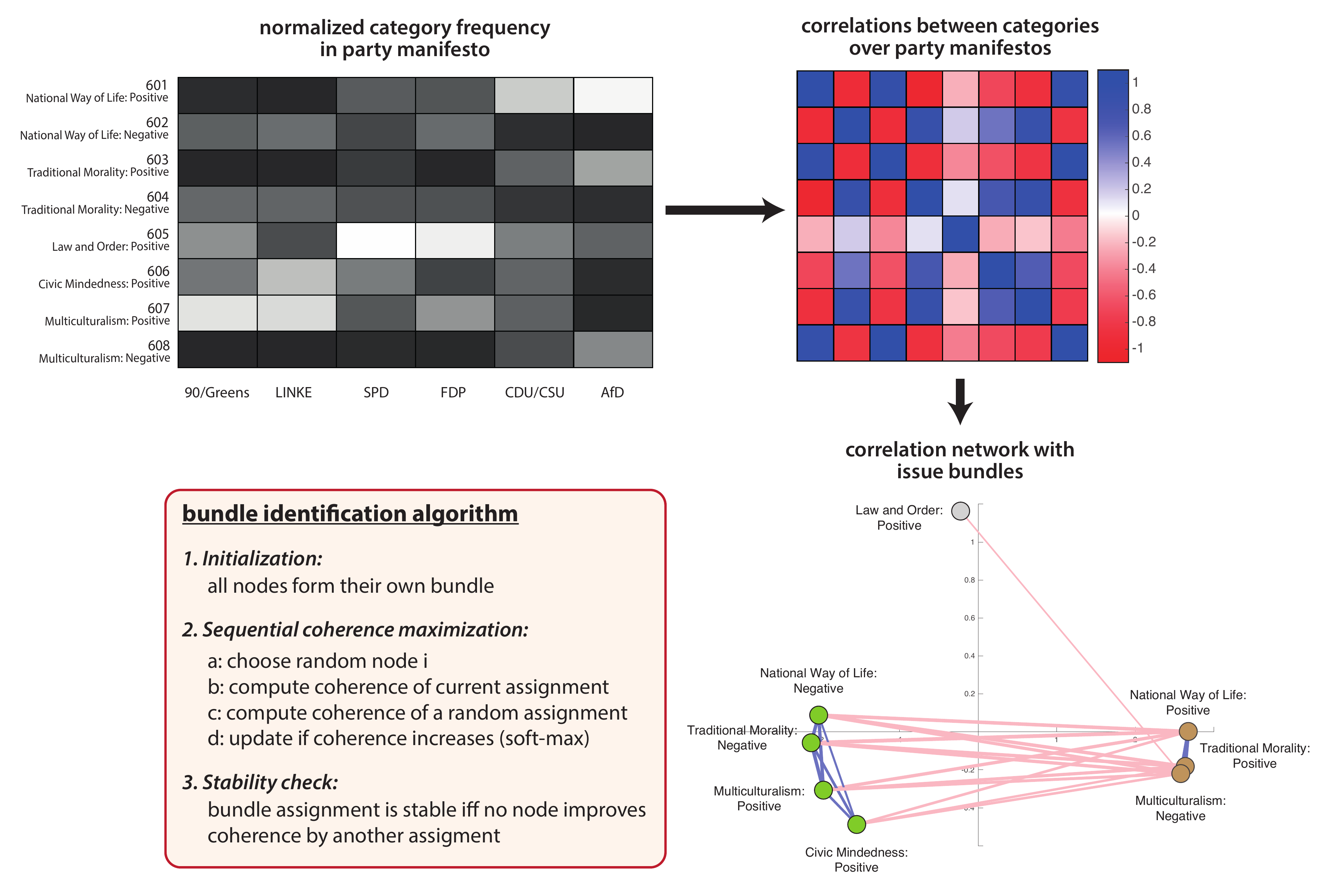}
	\caption{The construction of IICNs and the identification of issue bundles at the example of the cultural domain (601 -- 608) in the German 2017 elections. Starting from the frequency with which categories occur in the manifestos of the six parties, inter-issue correlations are computed ($R$) and visualized as a signed network. Bundles of correlated categories are clearly identifiable and the pseudo-code for bundle identification is provided in the red box.
	}
	\label{fig:exampleKUL2017}
\end{figure}

In order to identify bundles of issues on which party positions systematically differentiate we propose an algorithm for network-based clustering of correlational data. We consider the correlation matrix $R$ capturing pairwise correlations between categories or topics over a set of party manifestos (elements $r_{ij}$ correspond to the correlation coefficient between category/topic $i$ and $j$). $R$ can be seen to define a signed weighted network which we shall refer to as an inter-issue consistency network (IICN). This representation is inspired by recent connectionist models of attitude networks and belief systems constructed on the basis of multiple item surveys \citep{Dalege2016}. An example IICN is shown in Fig \ref{fig:exampleKUL2017} for the subset of categories in the cultural domain of the MARPOR data ("fabric of society") in the German 2017 election.

This example is chosen to illustrate what kind of information is captured with a IICN representation and what we mean by issue bundles. The normalized frequency with which the 8 associated categories occur in the manifestos for the 2017 German election is shown on the left of Fig. \ref{fig:exampleKUL2017}. We observe, for instance, in the first (601 - National Way of Life: Positive) and the third category (603 - Traditional Morality: Positive) that the Green Party and the Left score low on these two categories, whereas the CDU/CSU as well as the AfD score high. On the other hand, the picture is reversed with respect to the counter categories 602 (National Way of Life: Negative) and 604 (Traditional Morality: Negative). This means that parties which take a positive (negative) stance towards a "National Way of Life" also tend to be positive (negative) regarding "Traditional Morality“. Hence, these issues are ideologically bundled.

In the network representation we show only edges with a large absolute value of correlation. If the correlation is high and positive, the two issues contribute in a similar way to differentiating political positions meaning that parties emphasising one of them also emphasise the other (e.g. 601 and 603 or 602 and 604). If the correlation is strongly negative, one issue is emphasized by one subset of parties and rarely mentioned by the others, whereas the second issue is frequently mentioned by the latter subset and not by the former set of parties. Strong negative correlations are hence especially informative about the ideological lines on which parties take opposing stances. Low correlations indicate that there is no such clear pattern of distinction over the set of parties.

In order to automatically identify a unique bundle assignment over the IICN which best captures its modular bundle structure, we define a global coherence measure in analogy to the negative energy in physical spin systems, because we maximize the coherence, while energy is minimized in physics.  Let $(I,R)$ denote the IICN and let $x$ be a vector that assigns each node in $I$ to a certain class or bundle ($x: C \Rightarrow B$). The global coherence of bundle assignment $x$ on an IICN with interaction matrix $R$ is defined as:
\begin{equation}
	\gamma := \sum_{i} \gamma_i
\end{equation}
with individual node contributions given by
\begin{equation}
   \gamma_i = \sum_{j} r_{ij} \delta(x_i,x_j)  - r_{ij} (1- \delta(x_i,x_j) ) \;.
\end{equation}
The Kronecker $\delta(x_i,x_j)$ is one if $x_i=x_j$ and zero if they are assigned to a different bundle. For all neighbors of each node $i$ we hence check whether the nodes are assigned to the same bundle and to what extend this is consistent with the correlation in the data. If two positively correlated issues are assigned to the same bundle ($x_i = x_j$ and $r_{ij} > 0$) this is consistent and contributes positively to the coherence. If two positively correlated issues are assigned to different bundles there is a negative contribution. Two negatively correlated issue, one the other hand, are consistent with respect to a bundle assignment $x$ if they are in different bundles.
Hence, in a coherent partition $x$, positively correlated issues should be assigned to the same bundle, while negatively correlated bundles are assigned to different bundles.

The problem of bundle identification is closely related to the problem of community detection on unsigned networks and marked by a similar computational complexity. While a lot of research has been invested on understanding the subtleties of different algorithms for community detection \citep{Fortunato2010}, there is no well-established solution for signed networks. 

Here we propose a sequential algorithm to identify coherent bundle assignments. Initially ($t = 0$) each node is assigned to its own bundle. Then we iterate a sequential reassignment process which locally increases the coherence function. At each step a node $i$ is chosen at random and $\gamma_i(x_i)$ is computed. The current coherence $\gamma_i(x_i)$ is compared to the coherence of a randomly chosen alternative assignment $a$ for node $i$, $\gamma_i(a)$, by $\Delta \gamma_i = \gamma_i(a)) - \gamma_i(x_i)$. The alternative $a$ should be accepted if the coherence increases ($\Delta \gamma_i > 0$) which is realized by an update probability $p(x_i = a) = 1/(1 + exp(\beta \Delta \gamma_i )$ (soft-max). The parameter $\beta$ governs the exploration rate or temperature.
Following recent work to identify cohesive subsets in the context of opinion dynamics models \cite{Banisch2019}, we consider a bundle assignment $x$ as stable if no individual node can increase coherence $\gamma_i$ by an alternative assignment. Notice, however, that an IICN can possess more than one such equilibrium assignment and that the algorithm described here converges to only one of them. Therefore, for the identification of a "best" bundle assignment, the algorithm is repeated 100 times and the assignment with maximal global coherence $\gamma$ is selected out of the set of stable ones (local maxima of $\gamma$). 

\begin{figure}[ht]
	\centering
	\includegraphics[width=0.99\linewidth]{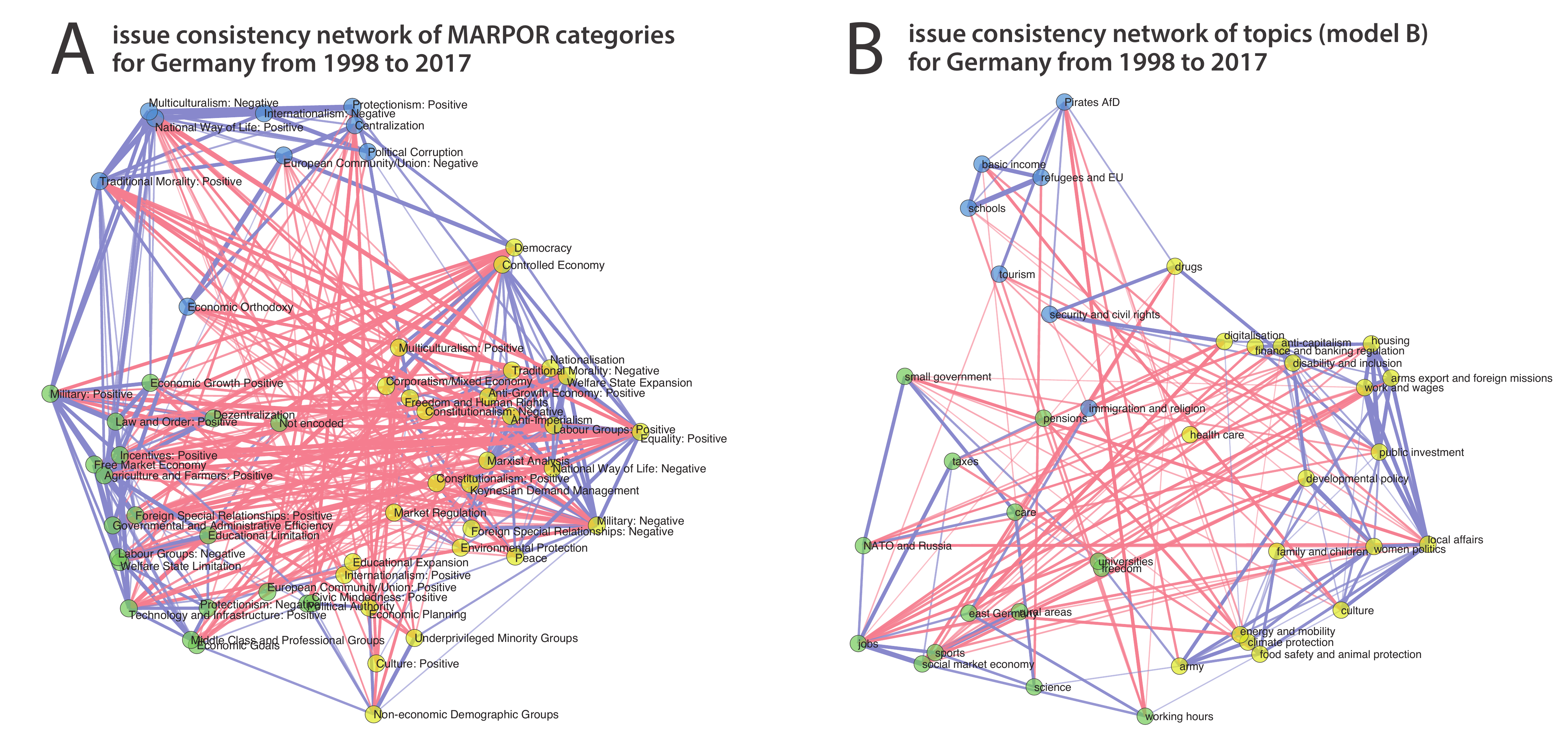}
	\caption{IICNs for categories (A) and topics (B) for Germany from 1998 to 2017. Edges between categories or topics are shown if their correlation is significant at the 5\% level ($p < 0.05$). Edge widths represent the strength of the correlations. The issue bundles that maximize coherence are marked by node color.
	}
	\label{fig:IICNall}
\end{figure}

For the analysis that follows we consider issue bundles derived for the entire time span from 1998 to 2017 as well as issues bundles derived single elections.
Fig. \ref{fig:IICNall} shows the IICN along with its bundles computed on the basis of the manifestos from 1998 to 2017. The left network is for category data (A) and the right on for topics (B).
These bundles are used in section \ref{sec:AfD}. 
In section \ref{sec:results_issue_bundles} we analyse the evolution of issue bundles by tracing changes from one election to the other.

\section{Results}
\label{sec:results}

\subsection{Robustness of the left-right dimension}

\begin{figure}[h!]
	\includegraphics[width=\linewidth]{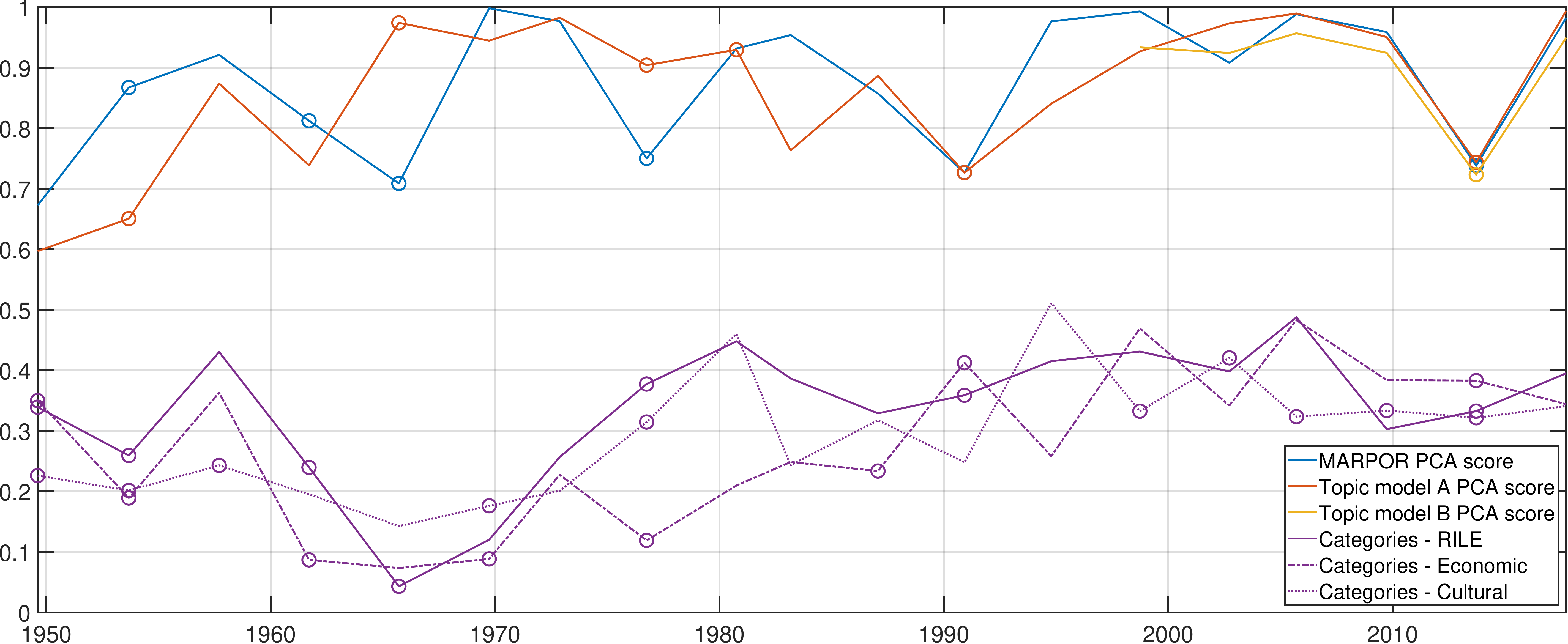}
	\caption{Blue, Orange, and yellow curves: Absolute values of correlations between the RILE score and the scores on principal components for single elections. Political spaces were estimated from MARPOR categories (orange), topic model A (yellow) and topic model B (purple). Circles indicate cases for which not the 1st component showed the strongest correlation. Purple curve: Maximal Correlation between the RILE axis and the economic and the cultural axis of \citep{Bakker2013}, respectively, and a principal component for single elections. The circle indicates if it is not the first principal component.}\label{fig:rile_correlations}
\end{figure}

A robust finding for using PCA to construct political spaces - be it for a single election or for a series of elections - is that very often the first principal component,  i.e. the direction with the largest variance in the issue space, show high correlations with the traditional left - right axis as it is represented by the RILE index. Fig.~\ref{fig:rile_correlations} illustrates this finding. The correlations between the RILE scores of the political parties and the scores of one of the principal components is always high, often larger than 0.9 for both the inductive political spaces constructed from the MARPOR categories and for the spaces estimated from the topic models. Lower, but still high correlations are observed for 1990 and 2013. In both years it is also not the first principal component that exhibits the highest correlations with the RILE index, which indicates that the left-right division was not the dominant difference at these elections. In 1990 the elections were dominated by questions related to the German re-unification and in 2013 the appearance of the AfD changed the political space. This case will be discussed in more detail below in \ref{sec:AfD}. The purple curves in Fig.~\ref{fig:rile_correlations} show correlations between the axes of the spaces itself. The vectors for the axis of the deductive spaces were built by setting the coordinates of the "right" categories on $+1$ and for the "left" categories on $-1$. However, this can be done only for the inductive space based on the MARPOR categories because they are projections from the same issue space as for the RILE space or economic-cultural space. For the spaces based on the topic models this is not possible because there the issue spaces are different. Interestingly, there are no systematic differences between the correlation with the RILE axis and the two other axes, indicating that they all are equally suited to describe the main political differences at single elections. In particular, we see no sign of a decline of the importance of the left-right distinction and a also no indication for an increasing importance of the cultural axis compared to the economic axis.    

\subsection{The appearance of the AfD in German political spaces}
\label{sec:AfD}

Given the robustness of the left-right axis in Germany, let us now look how the AfD appeared in these spaces. Fig.~\ref{fig:rile} shows the RILE index for the German parties contained in the Manifesto database. The party positions for most parties correspond roughly to the public perception. One could argue about relative ordering between SPD and Greens or the Christian Democrats and the Liberals on this axis, and there is a large literature about the validity of the RILE index, but this is not the focus of this paper. There is, however, also a significant difference to public perception: although already 2013 the AfD was widely perceived in Germany as a party right of the CDU/CSU it appeared in the middle of the RILE scale, to the left of the CDU/CSU. 

A first explanation can be gained by looking at the two-dimensional economic-cultural space of \cite{Bakker2013} in Fig.~\ref{fig:BakkerHobolt}. Here the the projection of the RILE space should be on the diagonal from the upper left corner to the lower right corner. Thus, although the AFD 2013 has a right-wing position on the economic axis, projected on the diagonal it would appear on the left of the CDU/CSU in 2013. The reason is the slightly ``leftish'' position on the cultural axis due to the very high count of more than 17 \% of the quasisentences counted in the category "Democracy" (202). This originated from the fact that on the one side the call for direct democracy based on the Swiss model is a central position in the program, but the program as whole was very short and thus covered only a small set of positions, which gave this demand this relative prominence. Obviously there are very different ideas about democracy (see for instance \citep{held2006models} for a discussion) and in particular there are doubts about the commitment of the AfD towards the liberal democracy, given the sympathy of their leader for proponents of an "illiberal democracy" such as Viktor Orban\footnote{See \url{https://www.afd.de/tag/orban/} for examples.}. Thus, using the category "Democracy" of the Manifesto coding scheme to estimate the position on a cultural liberal-authoritarian axis might in the case of the AfD actually conflate the distinction that one actually wants to measure. The latest coding scheme (version 5) of the Manifesto project reacted to such and similar problems in other categories by refining them: Now the Democracy category has 4 subcategories: Democracy General: Positive, Democracy General: Negative, Representative Democracy: Positive and Direct Democracy: Positive \citep{Manifesto2020-codebook}. Unfortunately, the refined coding scheme is only available for the 2017 elections and therefore we cannot test, how it would change the position of the AfD in a refined space. We can, however, study how the AfD gets positioned in the inductive political spaces.     

\begin{figure}[th]
	\includegraphics[width=\linewidth]{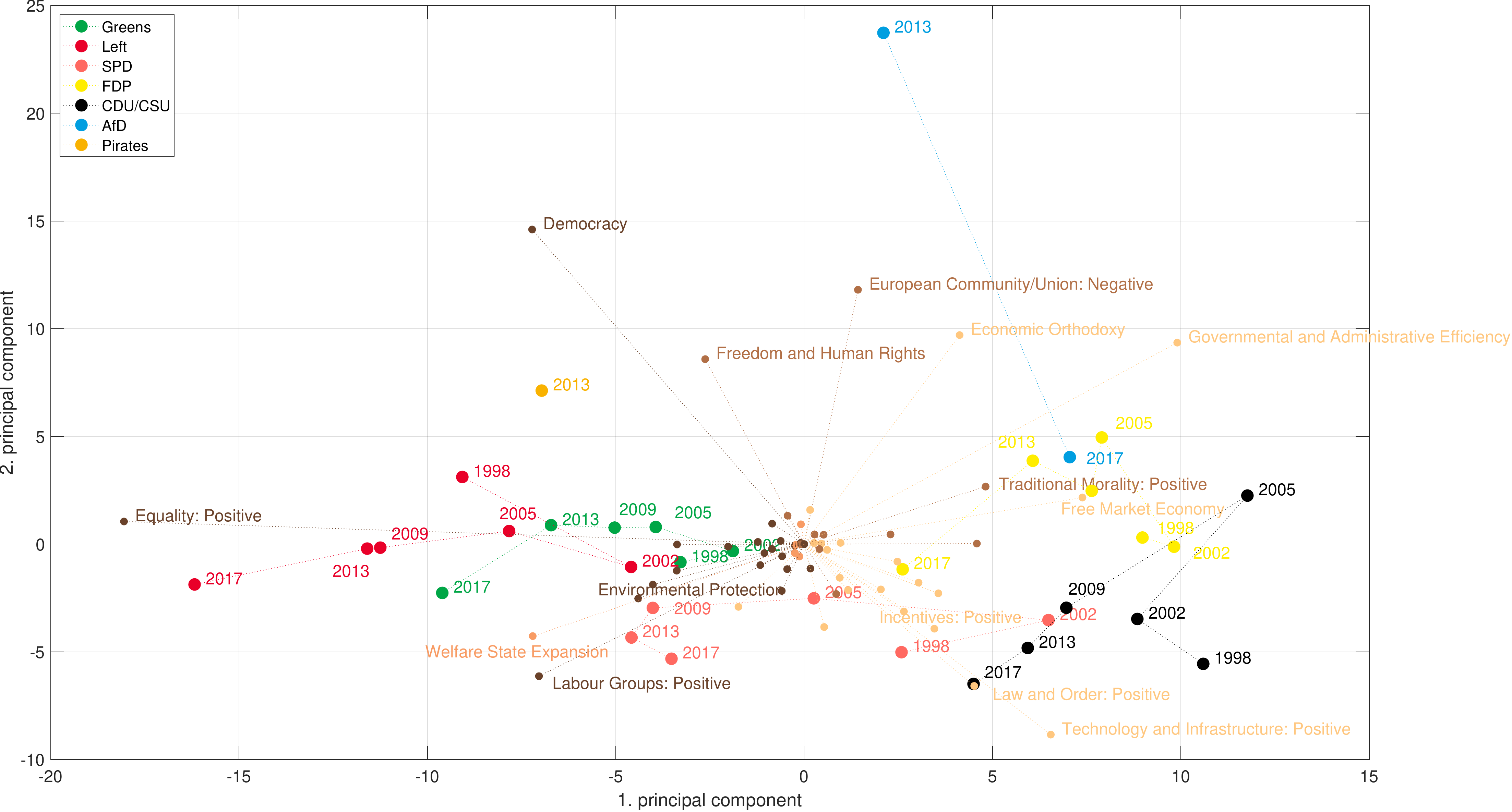}
	\caption{First two principal components for the political estimated from the MARPOR categories for Germany 1998 - 2017. For the party positions it is identical to Fig.~\ref{fig:pca_1998_2017_1_2} A. Here we show additional the directions of the issues in this space. Note that only the relative length of the issue vectors are meaningful. For better readability, the labels are only shown for issues with sufficiently large projections in this space. The colors (different copper tones) indicate to which of the 3 issue bundles the issues belong.}  \label{fig:MARPOR_PCA_1_2}
\end{figure}

Fig.~\ref{fig:MARPOR_PCA_1_2} shows the positions of the parties in the space spanned by the first two principal components estimated from the counts on the MARPOR categories. It is the same space as shown in Fig.~\ref{fig:pca_1998_2017_1_2} A. Additionally, it also shows how the different issues project in this space. The vectors were scaled for better visibility, thus only the relative length and the direction of these vectors have meaning. The colors indicate to which issue bundle the categories belong (see Fig. \ref{fig:IICNall}). For this representation the issue bundles were estimated for the whole time period, thus for computing the PCA and the issue bundles the same correlation matrix was utilized.

Looking at Fig.~\ref{fig:MARPOR_PCA_1_2} one can make a few immediate observations: The first principal component orders the parties on a left - right axis, as we would expect from Fig.~\ref{fig:rile_correlations}. Second, projected on this axis, also here the AfD would appear in the middle, on the left of the CDU/CSU and moves to the right only in 2017. But, most striking, the second dimension is almost exclusively spanned by the Afd. In comparison to the AfD there are only small movements of the other parties in this direction (perhaps except the Pirates). Moreover, we see that the category that points directly to the AfD, is "European Union: Negative", which reflects the relative prominence of this category in the AfD 2013 program, however with "Freedom and Human Rights" and "Economic Orthodoxy" close-by. Interestingly, next to "Freedom and Human Rights" is "Democracy" which first points already to the left and second is not in the same issue bundle as the other three categories, which indicates that is has stronger ties with more leftish issues, than with the issues that are prominent in the AfD program, despite the fact that it is the most frequent category in the AfD program 2013. 

\begin{figure}[th]
	\includegraphics[width=\linewidth]{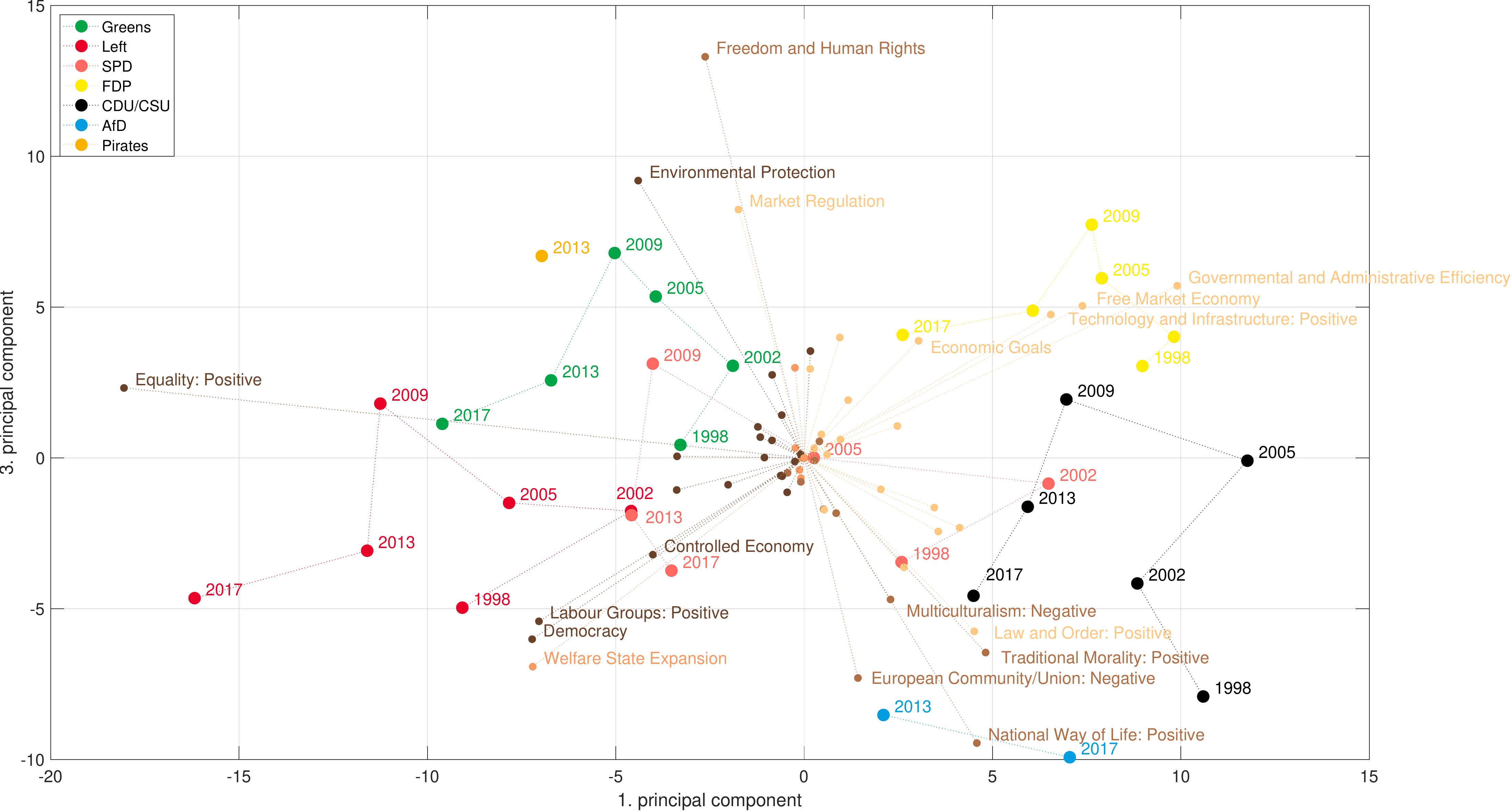}
	\caption{Same as in Fig.~\ref{fig:MARPOR_PCA_1_2} but now for the 1st and the 3rd principal component.}  \label{fig:MARPOR_PCA_1_3}
\end{figure}

Does this mean that the economic-cultural space was the wrong space to look at for the 2013 elections? If one is interested in the specifics of the AfD, yes. But, let us look at the space spanned by the first and third principal component Fig.~\ref{fig:MARPOR_PCA_1_3}. By rotating it by approximately $45$ degrees we recover a kind of economic-cultural space. We see categories such as "Welfare state expansion", "Equality", "Controlled economy" or "Labour groups: positive" on the economic left side and "Free market economy" on the economic right side, while "Environmental Protection" is prominent on the cultural left and "Multiculturalism: Negative", "Traditional Morality: Positive", "National Way of Life: Positive" and "Law and Order: Positive" appear as cultural right. There are, however, also differences to the axis used by \citet{Bakker2013}: Democracy appears on the economic left axis, Market regulation appears on the "cultural" axis and "Equality: Positive" is located in between the economic and the cultural axis. While some of this might make sense intuitively, such as for "Equality: positive", because it includes not only inequalities due to class, but also references to end racial or sexual discrimination, it is less clear for the other categories. Regarding the positioning of the AfD, we see that it now appears on the "cultural right" axis and on an economically more left position compared to the deductive space in Fig.~\ref{fig:BakkerHobolt}. Note that the problem that we identified exemplarily for the "Democracy" category still persists in this space and leads to the surprisingly leftish position on the  "economic" axis in this space. 

\begin{figure}[h!]
	\includegraphics[width=\linewidth]{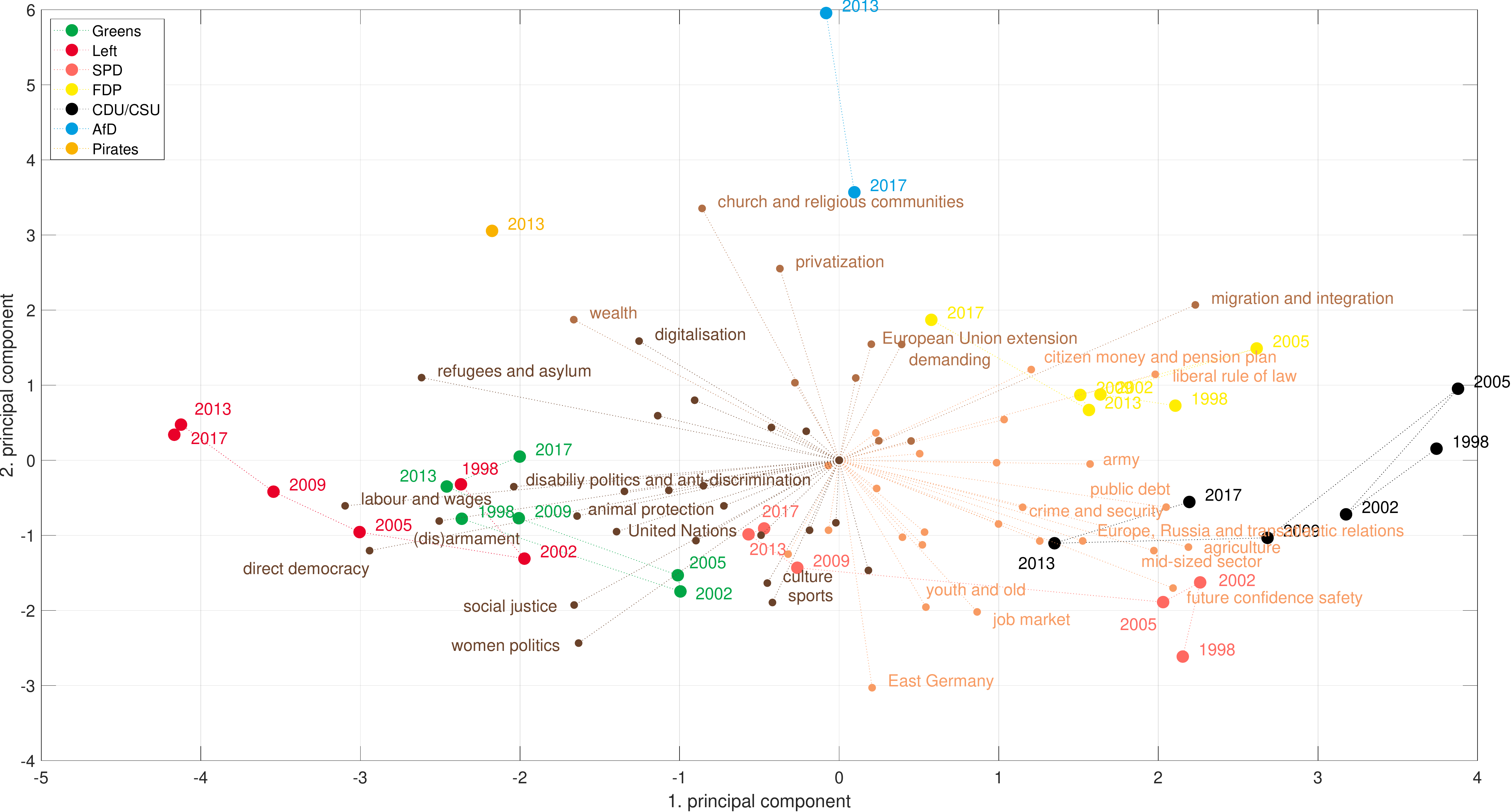}
	\caption{Same as in Fig.~\ref{fig:MARPOR_PCA_1_2} but now using the topic probabilities from topic model A.}  \label{fig:topic_A_PCA_1_2}
\end{figure}

\begin{figure}[h!]
	\includegraphics[width=\linewidth]{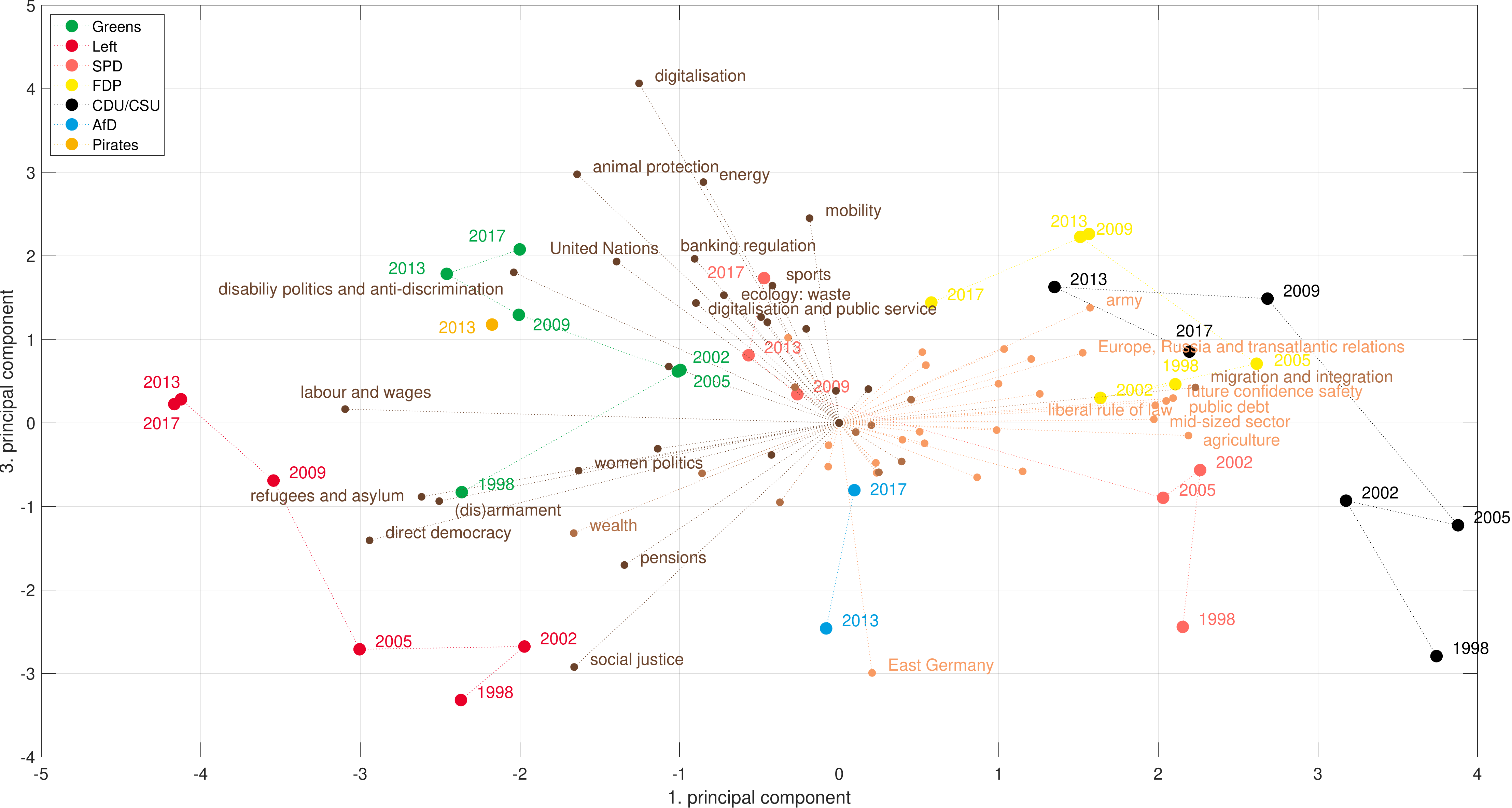}
	\caption{Same as in Fig.~\ref{fig:topic_A_PCA_1_2} but now for the 1st and the 3rd principal component.}  \label{fig:topic_A_PCA_1_3}
\end{figure}

Therefore, in a next step, we look at the political spaces constructed from the topic models. There, we do not only not specify the axes in advance, but we also learn the categories from the data in an unsupervised fashion. Fig.~\ref{fig:topic_A_PCA_1_2} and Fig.~\ref{fig:topic_A_PCA_1_3} show plots corresponding to Figs.~\ref{fig:MARPOR_PCA_1_2} and \ref{fig:MARPOR_PCA_1_3}, respectively, for topic model A (see \ref{meth:topic_models}). Although there are differences between the different political spaces, there are also remarkable similarities. The AfD spans their own dimension. Also with the topic models it is the 2nd principal component\footnote{This depends on the details of the construction of the political space. Doing it only for the 2013 election it becomes the 1st principal component in both cases either using the MARPOR categories or the topic model (see also Fig.~\ref{fig:rile_correlations}).}. Relevant topics for this dimension are here the ones called "church and religious communities" which contains the Islam, "privatization", and "migration and integration", which reflects one of the main programmatic topics of the AfD. Note that the topic model has two topics containing migration: topic 52 labeled as "refugees and asylum" and also stongly articulated by the Greens and the Left and topic 40 "migrationa and integration" that is more prevalent with the CDU/CSU and contains words related to integration and corresponding problems such as "Parallelgesellschaften" (parallel societies). The positioning of the parties is similar in the two political spaces, in particular in the space spanned by the first two components. There is also a clear correspondence between the issue bundles derived from the MARPOR categories and the ones derived from the topic models (Fig.~\ref{fig:IICNall} in terms of their location in the political space (see Figs \ref{fig:MARPOR_PCA_1_2} and \ref{fig:topic_A_PCA_1_2}). There is a "leftish" bundle, a "rightish" bundle and one related to the dimension spanned by the AfD.

Looking at Fig.~\ref{fig:topic_A_PCA_1_3} it is even less possible to make a clear distinction between an economic and a cultural axis than in Fig.~\ref{fig:MARPOR_PCA_1_3}. While there are clearly economic topics such as "labour and wages" or "housing" on the left side and "public debt" or "taxes" on the right side, they are not clearly divided from non-economic topics such as (dis)armament or "women politics" on the left side and "liberal rule of law" or "Europe, Russia and transatlantic relations" on the right side. Thus, we do not observe two clearly distinguishable axes with a different political semantics beyond a rough left-right division in this space. This is also reflected by the fact that the space is mainly populated by topics from two issue bundles and the topics from the third one ("wealth" and "migration and integration") define not an own region in this space. Thus, using topic models does not resolve the problems that appear with some of the MARPOR categories, in particular, parties that articulate new concepts. Although the topic model identifies new or specific topics, such as "housing" or "East Germany", and although it gets more specific (for instance environmental protection gives rise to at least 4 topics), it has the problem that it often does not resolve the polarity of the topic. That means that pro and contra positions appear in the same topic. For instance, we observed that in the "refugees and asylum" topic (52) or in the Euro topic (38). Therefore, in particular controversial topics for which opposing positions were articulated by the different parties these positions do not appear in the political space on opposing ends of an axis (see for instance the topic "refugees and asylum" in \ref{fig:topic_A_PCA_1_3} and the positions of the Left, Greens and the AfD). Despite these limitations, topic models can be used to construct meaningful political spaces (see Fig. \ref{fig:pca_1998_2017_1_2}).

What do these political spaces tell us about a potential representation gap? We have not seen a gap in the one-dimensional left-right spaces. Even though we could observe some movement to the left between 2002 and 2013, the AfD seemed to appear in the middle. However, looking at additional dimensions we found that (1.) the AfD created a new dimension of the political space by its own combination of issues and (2.) that in the two-dimensional space which resembles to some extent the economic-cultural space of \citep{Bakker2013} (Fig.~\ref{fig:MARPOR_PCA_1_3}) the AfD appeared at cultural right-wing positions from which both the SPD and the CDU have moved away after 1998. In fact, we see a movement of the CDU along the cultural axis\footnote{The downward movement after 2009, however, is not in agreement with the narrative of a liberalisation of the CDU under the chancellorship of Angela Merkel.} and a movement to the economic right of the Schröder SPD from 1998 to 2002, which is then followed by leftward movement in the subsequent years. Thus, we find some indications for a representation gap if we look at a space fixed between 1998 and 2017, but also the strong signal of a new dimension that needs further analysis. A new dimension means that there are new correlation patterns between the issues and we will use the concept of issue bundles (section \ref{method:issue_bundles}) to study the evolution of the issue correlations between the different elections. 

\subsection{Temporal evolution of issue bundles}
\label{sec:results_issue_bundles}

Issue bundles -- derived in a purely inductive way -- can provide a complementary perspective on questions related to the reconfiguration of the political space. In particular, we may ask if there are bundles that persist over the time span from 1990 to 2017 and if the entry of the populist AfD introduces a new pole.
For this purpose, we compute IICNs from the MARPOR data for each election from 1990 to 2017 and identify the respective issue bundles by the algorithm described in Sec. \ref{method:issue_bundles}.
We can then relate the bundles from one election to those of the next one and analyse their transformations. 
This requires to characterize and identify bundles over time which we did by calculating their overlap with the sets of categories that define poles of the RILE axis and the economic and cultural axis, respectively, using the Jaccard similarity coefficient. Once this is established, we can trace how new bundles emerging in an election draw from the different issue bundles of the previous one.

\begin{figure}[t!]
	\centering
	\includegraphics[width=0.89\linewidth]{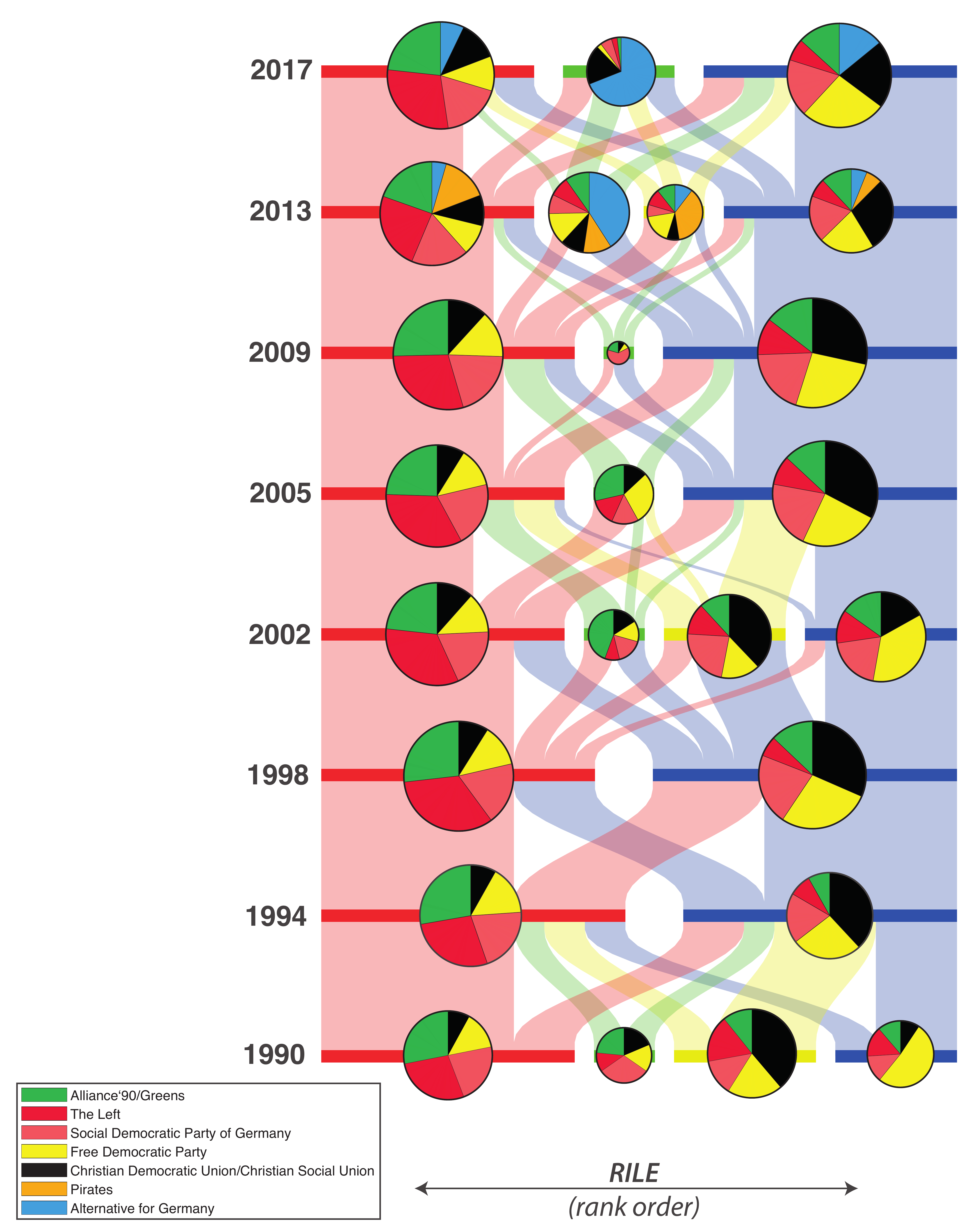}
	\caption{Time evolution of issue bundles (categories) from 1990 to 2017. Bundles are represented by horizontal bars and ordered according to their overlap with the RILE index. For each bundle the coverage by the different parties is shown as a pie chart. 
	}
	\label{fig:draftverti01}
\end{figure}

Fig. \ref{fig:draftverti01} shows the temporal evolution of issue bundles during the last eight elections in Germany. 
Each bundle is represented by a horizontal bar with a size proportional to the number of categories it comprises along with a pie chart that shows how different parties cover the respective subsets of categories.
Notice that for each election the bundles are ordered according to their overlap with the RILE categories.
The flow chart from one election to the next shows how many categories remain in the respective bundle and how many categories are re-assigned into a new or an alternative bundle.

We observe that the number of stable issue bundles ranges from two in 1994 and 1998 to four in 1990, 2002 and 2013.
We also observe that the issue bundle to the left is relatively stable over time and mainly covered by the Green Party, the Left and, to a minor extend by the Social Democrats.
The core categories that remain within this bundle through the entire time span are "Military: Negative", "Equality: Positive", "National Way of Life: Negative", "Labour Groups: Positive“.
There are slight deviations in 2002 where a "green bundle" emerges (Greens 2002) and in 2005 where the Green Party and the FDP occupy a small bundle in the center.
There is slightly more variation to the right of the spectrum mainly due to the fact that in some elections (1990 and 2002) the center-right cluster splits into a bundle covered mainly by CDU/CSU and another one by the FDP. 
According to the RILE order, the set of categories consistently promoted by the FDP is to the right of the CDU/CSU cluster.
All in all, we observe a clear signature of a leftish bundle occupied mainly by the Left and the Greens and a right-leaning bundles of categories occupied by the FDP and the CDU/CSU.
In this analysis the German SPD is shown to promote positions from both sides.

The two new parties that entered the Bundestag in 2013 -- the Pirates and the AfD -- affect the IICNs and show up in two independent bundles in that year.
Noteworthy, the "Pirate bundle" is right to the "AfD bundle" according to the RILE classification which hints at the fact that the RILE index might be inappropriate to capture the new populist right.
Indeed, this classification is due to the category "Freedom and Human Rights" being part of the "Pirates bundle" and considered to be a "right" category in the RILE, while on the other side "Democracy" is part of the "AfD bundle" and classified as "Left" by the RILE.  The issue bundle strongly associated to the AfD remains stable in 2017 where it still appears in between the left bundle and what we refer to as the center-right bundle.
Noteworthy, there are significant flows from all bundles in 2013 to the 2017 AfD bundle.

\begin{figure}[t!]
	\centering
	\includegraphics[width=0.89\linewidth]{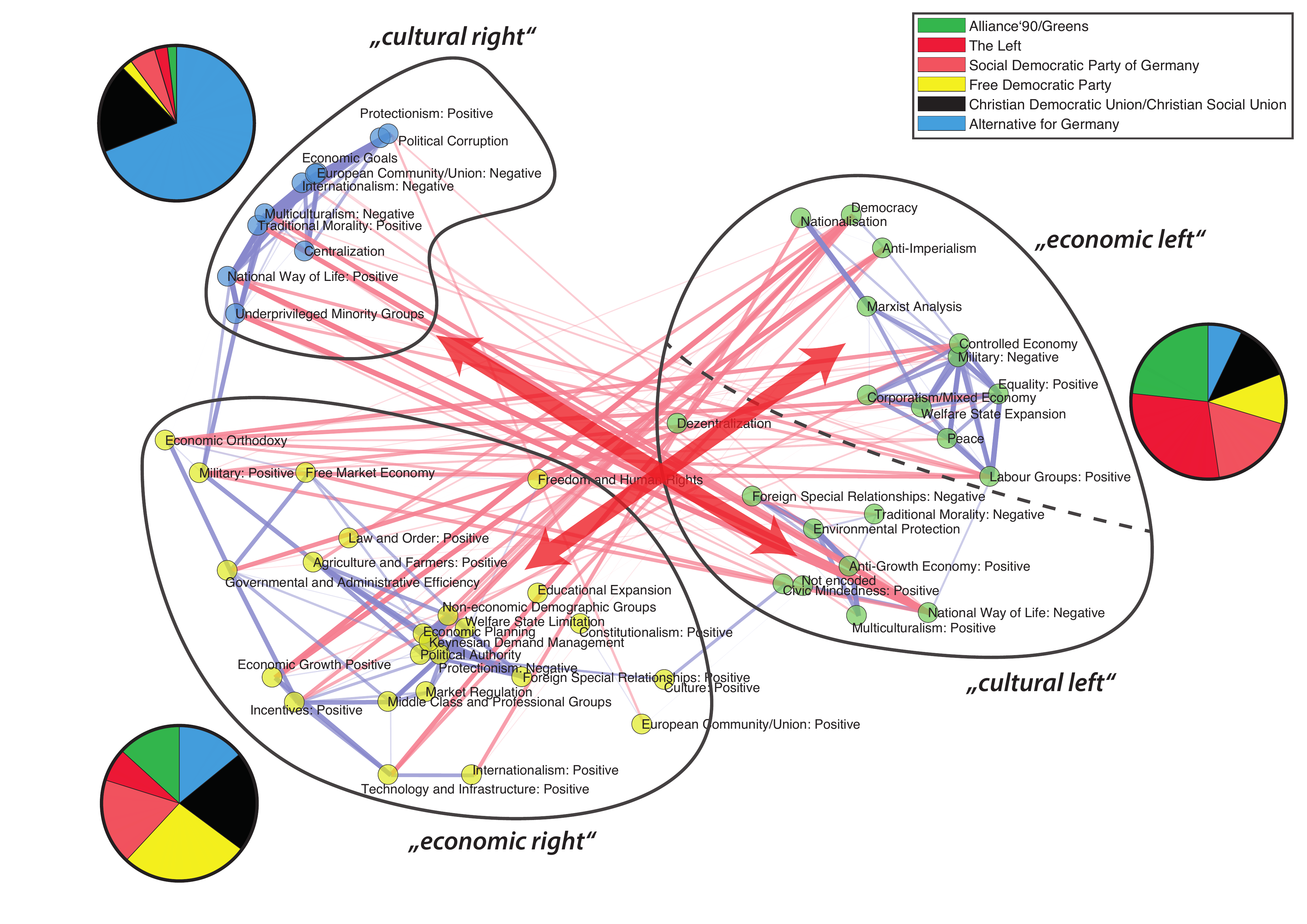}
	\caption{IICN for the German elections in 2017. Blue links encode strong positive correlations and red links negative ones. Only correlations significant at the 10\%-level ($p < 0.1$) are shown and the thickness of an edge reflects significance and strength of the correlation. The three subsets of categories corresponding to the three different issue bundles are highlighted and the coverage by the different parties is shown. 
	}
	\label{fig:catsbundles2017ai}
\end{figure}

A closer look into the bundle structure of the 2017 IICN is provided in Fig. \ref{fig:catsbundles2017ai}.
It reveals that the AfD 2017 occupies a mixture of categories related to the "cultural right" as well as to "anti-globalization" marked by categories such as "European Community/Union: Negative", "Internationalism: Negative" and "Multiculturalism: Negative". 
This cluster also contains economic issues such as "Protectionism: Positive" or "Economic Goals", but the coherence corresponds rather to a cultural logic than to an economic logic.
On the other hand, we find a bundle that could be referred to as "economic right". This cluster comprises pro globalization categories such as "Protectionism: Negative", "European Community/Union: Positive" and "Internationalism: Positive".
This supports the hypothesis that topics related to globalization draw a new line of distinction on the right of the political spectrum.
Thirdly, there is a large bundle of issues assigned to the left bundle mainly covered by the Left and the Greens.

In the network we emphasize those connections for which the inter--issue correlation is significant and large. 
In this way two axes of strongly negative associations become visible.
On the one hand, we find the more classical axis of distinction on economic issues between the "economic right" and the upper part of the left bundle labeled with "economic left".
On the other hand, several categories in the new issue bundle promoted by the AfD have a strong negative correlation with issues from a "cultural left" and "National Way of Life: Negative" in particular.
In this way, IICNs render visible how the new cultural divide \cite{Bornschier2010} associated to the emergence of right populism is reflected in the programs of political parties.
However, our analysis also suggests that -- at least in Germany -- the classical economic divide remains an equally important axis of political distinction.

\section{Discussion}
\label{sec:discussion}
We analyzed the German political space for the elections in the last decades and compared deductive and inductive spaces. We constructed the inductive political spaces by doing principal component analysis on the issue probabilities or the logarithms of the topic probabilities, respectively. The first principal components indicate the directions in the issue space on which the party positions differ most. We found that the dimension on which party positions differ most in the inductive spaces, the first principal component, can still be characterized as a left-right dimension. However, if one does the analysis for each election separately one sees, that there are changes over time. This is reflected in the changing content of the "leftish" issue bundle while having a stable core (Fig.~\ref{fig:catsbundles2017ai}). Moreover, the political spaces constructed from the topic models and those constructed from the Manifesto categories showed a similar topology for the different parties, despite that fact that they were constructed from quite different issue spaces. Thus, we demonstrated that it is possible to construct meaningful political spaces in a completely unsupervised fashion using topic models. 

Studying the appearance of the AfD in these political spaces, we found some indications for a representation gap, but only in the inductive spaces.
We have found that the inductive spaces spanned by the first and the third principal component (Figs.~\ref{fig:MARPOR_PCA_1_3} and \ref{fig:topic_A_PCA_1_3}) are similar to the deductive economic-cultural space shown in Fig.~\ref{fig:BakkerHobolt} with the first principal component being more economic and the third more cultural.  However, in the inductive spaces one can observe between 1998 and 2009 an upward movement of all parties, i.e. on the dimension containing issues such as "Freedom and Human Rights", "Environmental protection" or "Market Regulation" an topics such as "digitalisation", "animal protection" or "banking regulation" leaving a gap in the lower part of the space that can be seen as an indicator for a representation gap. It is then also this region where the AfD is appeared.  However, it has to be noted that these spaces were created including the AfD manifestos and therefore the "gap" in this space would have not occurred in this way in a political space created from the party manifestos without 2013 and 2017.
Our main finding, however, was that the main effect of the appearance of the AfD was that it occupied a new dimension of the political space, characterized by a new issue bundle combining issues and topics such as "European Union" or "Immigration" with issues, such as "Democracy" or "Protectionism: positive" that appeared on the left in the RILE scale.

In order to analyze these issue combinations in more detail we operationalize the concept of issue bundles \citep{Daeubler2017estimating} to be inductively inferred from the data. Inter-issue consistency networks (IICNs) represent the correlations between issues over a set of manifestos as a signed weighted graph and reveal systematic patterns of consistency across issues given a set of party manifestos. We developed an algorithm to uniquely identify issue bundles defined as stable subsets of consistent categories. We estimated issue bundles for each election and followed their evolution over time. This analysis also reflected the transformation of the AfD from 2013 to 2017, from the Anti-Euro party with a strong emphasis on economic arguments (reflected by the MARPOR category "Economic Orthodoxy) to a national-conservative anti-immigration party. 
Identifying issue bundles in inter-issue consistency networks (IICN) turned out to be a very instructive for understanding the emergence of the new dimension in political space, because they correspond to specific regions in political spaces. Therefore, bundle structure of IICNs provides a complementary perspective on political spaces which is more sensitive to information related to the meaning of political axes. More generally, this method seems to be promising for the further exploration of the link between natural language processing techniques on the one side and conceptual spaces on the other side, because the political spaces considered in this paper can technically be seen also as document embeddings in a semantic space.

Issue bundles are closely related to deductive scales such as the RILE index:  they both define subsets of categories that point into the same direction. The fact that the AfD bundle inferred from the data cuts across the left and right set of categories in the RILE seems to suggest that the logic behind the RILE is different from the logic behind the political goals promoted by the AfD. This is reflected in the emergence of the issue bundle denoted as "cultural right" in Fig.~\ref{fig:catsbundles2017ai}. Containing issues such as "Multiculturalism: Negative", "National way of Life: Positive", but also "European Community/Union: Negative" it corresponds to one pole of the new cleavage called  "traditionalist-communitarian" by \citet{Bornschier2010}. The inductive approach of issue bundle identification will be helpful for further research on the rise of populism in Western Europe as it enables a systematic characterization of "populist profiles" in a cross-national perspective.

There are also clear limitations of the presented approach. First, as already discussed in the introduction, we only study the party positions (the supply side), but not the positions of the voters (the demand side). Moreover, by looking only at the manifestos we do not consider which positions the political parties express during their election campaigns in speeches, public debates, advertisements or on social media outlets. In particular, for a populist party such as the AfD this could be very different and the positions stated in the electoral manifesto might be more moderate in comparison.

Second, we used principal component analysis to construct the inductive political spaces. There are many possible alternatives for doing the dimension reduction from the issue space to the political space, such as, for instance, multidimensional scaling. This term comprises methods that try to map points from a high-dimensional space in a low-dimensional space by preserving the distances between the points as good as possible and it is usually formulated as an optimization problem. \cite{Kriesi2006} is an example for its use for constructing political spaces. A similar method based on information theoretic principles, which is in particular popular in machine learning, is t-distributed stochastic neighbor embedding (t-SNE) \citep{Maaten2008}. The main problem with these methods in our context is, that they do not provide a clear mathematical interpretation of the axes of the resulting spaces, while the principal component analysis generates uncorrelated axis. We also desisted from rotating the axis as it is done in factor analysis, because we think that it is useful to ask for the axis that shows the largest variation of political positions, as the first principal component is doing.

There is another aspect that comes into play in the case of political spaces: If people use the positions of political parties or candidates in political spaces to make their voting decisions, how do these people actually perform the dimension reduction?  What is the metrics they use? What are the spaces that one had to consider for modeling opinion dynamics in such spaces by applying ``bounded confidence'' models such as the Hegselmann and Krause model \citep{Hegselmann2002opinion} or the interacting argument model in \cite{Banisch2021}? We will address these questions in our future research.

\section*{Funding}
This paper is part of the ODYCCEUS project. This project has received funding from the European Union’s Horizon 2020 research and innovation programme under grant agreement No 732942. 

\section*{Acknowledgments}
We would like to thank Armin Pournaki for helpful comments and suggestions. 

\section*{Data Availability Statement}
The dataset analyzed for this study can be found in the web page of the manifesto project \url{https://manifesto-project.wzb.eu/}.

\bibliographystyle{unsrtnat}
\bibliography{reconfig}  
\vfill
\begin{appendix}
\section{Topic models}
\begin{landscape}

\begin{table}[t]
	\centering
	\begin{tabular}{|l|c|c|c|} \hline
	number & $\lambda = 1$ & $\lambda = 0.3$ & English label \\ \hline 
1	&     stärken wichtig setzen &	    wichtig stärken leisten &	  	strengthen \\
2	&    fordern insbesondere Förderung & fordern staatlich Verbesserung &	  demanding \\
3	&    Deutschland setzen deutsch	&    Deutschland frei Demokrat deutsch	&    	Germany \\
4	  &  schaffen stellen stehen	 &   schaffen GRÜNE grün	  &   create \\
5	  &  Deutschland stark Land	&    starke erfolgreich Regel	&    Germany and Europe \\
6	  &  fördern langfristig zielen	&    Kernenergie Verwendung langfristig  &	 nuclear power \\
7	  &  Gesellschaft Leben leben	  &  Gesellschaft Teilhabe Leben	  &   diversity \\
8	  &  Zukunft Politik Deutschland	&    CDU CSU mein Land sozial Marktwirtschaft	 &  future confidence safety \\
9	  &  sozial soziale Politik	&    PDS Sozialstaat sozial	&    	social justice \\
10	 &   Ausbau Maßnahme Entwicklung &	  Bundesbahn Ausbau andererseits & 	  railroad \\
11	 &   Kommune Bund Land	 &   Kommune Bund Bund Land  &	local affairs  \\
12	 &   öffentlich Unternehmen Investition  &	  Digitalisierung Investition digital &	digitalisation and public service \\
13	&    Staat bürgern Freiheit	     & liberal bürgern Liberale &	liberal rule of law \\
14	&    alt Jugendliche jung	    & Jugendliche junge alt &	youth and old \\
15	&    europäisch EU national	&    Türkei EU europäisch &	European Union extension \\
16	&    fordern öffentlich DDR	 & Bürgerin DDR BRD &	DDR BRD  \\
17	&    Leistung Beitrag erhalten	     & Leistung Bürgergeld Altersvorsorge &	Citizen money and pension plan  \\
18	&    Bundesregierung Sozialdemokrat SPD  & 	   Sozialdemokrat sozialdemokratisch  &	federalism \\
    &   &  geführt modern Industriegesellschaft & \\
19	&    Zusammenarbeit Staat Europa  &	    Berlin Deutschlandpolitik Beziehung	&	policy of détente \\
20	&    Unternehmen Mittelstand Arbeitsplätze	  &  Mittelstand mittler Unternehmen  &	mid-sized sector \\
& & geführt modern Industriegesellschaft & \\
21	&    Staat Investition schaffen	 &   Schulden Neuverschuldung ausgeben	& public debt  \\
22	&	    Privatisierung Unternehmen öffentlich	&	    Privatisierung privatisieren Subvention	&	privatization	\\
23	&	    fordern Eigentum wirtschaftlich	&	    Eigentum Sozialisierung	&	property  (consequences of	\\
    &                                       &                               &   war and displacement)  \\
24	&	    Arbeitnehmer Unternehmen Betrieb	&	    Arbeitnehmer Mitbestimmung betrieblich	&	employee participation	\\
25	&	    ökologisch Umwelt Wirtschaft	&	    Umwelt Umweltpolitik Umweltschutz	&	environmental protection	\\
26	&	    Demokratie Partei Politik	&	    Demokratie Volksentscheid Einmischen	&	direct democracy	\\
27	&	    Partei politisch Kirche	&	    Kirche Heimatvertriebenen Abgeordnete	&	church and religious communities	\\
28	&	    Arbeit Beschäftigte LINKE	&	    LINKE Beschäftigte Leiharbeit	&	labour and wages	\\
29	&	    Welt Entwicklung Land	&	    Entwicklungsländer Entwicklungsländern 	&	developmental policy	\\
& & Entwicklungspolitik & \\ \hline
\end{tabular}
\caption{First 30 topics of topic model A trained on German manifestos from 1949 to 2017. Shown are the 3 most relevant words for $\lambda =0$ and $\lambda =0.3$ (see Eq.~\ref{eq:relevance}) and the assigned English labels.}
\label{tab:topic_model_A1}
\end{table}

\begin{table}[t]
	\centering
	\begin{tabular}{|l|c|c|c|} \hline
		number & $\lambda = 1$ & $\lambda = 0.3$ & English label \\ \hline 
30	&	    Arbeit Arbeitsmarkt Arbeitslosigkeit	&	    Arbeitsmarkt Arbeitslose Langzeitarbeitslose	&	job market	\\
31	&	    Behinderung Diskriminierung Schutz	&	    Behinderung behindert Barrierefreiheit	&	disabiliy politics and \\ & & & anti-discrimination	\\
32	&	    international Vereinte Nation setzen	&	    Vereinte Nation UNO Menschenrechte	&	United Nations	\\
33	&	    Ostdeutschland Land Region	&	    Ostdeutschland ostdeutsch Aufbau Ost	&	East Germany	\\
34	&	    Leben Schutz Hilfe	&	    Cannabis Droge straffrei	&	drugs, abortion and birth control	\\
35	&	    Europa Welt gemeinsam	&	    Russland liberal Außenpolitik transatlantischen	&	Europe, Russia and \\ & & & transatlantic relations	\\
36	&	    Volk Politik Deutschland	&	    CDU deutsch Volk Volk	&	Germany reunification	\\
37	&	    Prozent steigen Million	&	    Prozent DM Vermögen	&	wealth	\\
38	&	    Europa europäisch EU	&	    Europa europäisch EU	&	Euro	\\
39	&	    Unternehmen Banken Risiko	&	    Banken Aufsicht Bankenaufsicht	&	banking regulation	\\
40	&	    Deutschland Integration Zuwanderung	&	    Zuwanderung Integration Einbürgerung	&	migration and integration	\\
41	&	    Landwirtschaft ländlich Raum Landwirt	&	    Landwirtschaft Landwirt ländlich Raum	&	agriculture	\\
42	&	    Rente alt Rentenversicherung	&	    Rente Rentenversicherung Alterssicherung	&	pensions	\\
43	&	    Kultur kulturell fördern	&	    Kultur Kunst Kultur Kunst	&	culture	\\
44	&	    Datum Internet Datenschutz	&	    Datum Piratenpartei Datenschutz	&	digitalisation	\\
45	&	    Bildung Kind Schule	&	    Schule schulen Schüler	&	school	\\
46	&	    Bundeswehr Einsatz	&	    Bundeswehr Wehrpflicht	&	army	\\
47	&	    Sport Engagement fördern	&	    Sport Rechtsextremismus Antisemitismus	&	sports	\\
48	&	    Unternehmen Steuerreform steuerlich	&	    Steuerreform Einkommensteuer Erbschaftsteuer	&	taxes	\\
49	&	    Frau Mann Frau Mann	&	    Frau Mann Frau Mann	&	women politics	\\
50	&	    Kind Familie Eltern	&	    Kind Familie Eltern	&	family and children	\\
51	&	    Sicherheit Polizei Schutz	&	    Polizei Kriminalität Sicherheitsbehörden	&	crime and security	\\
52	&	    Flüchtling Schutz abschaffen	&	    Flüchtling Abschiebung Flüchtlingspolitik	&	refugees and asylum	\\
53	&	    Wohnung Stadt sozial	&	    Wohnung Wohnraum Mieterinnen Mieter	&	housing	\\
54	&	    militärisch Europa Abrüstung	&	    Abrüstung militärisch Waffe	&	(dis)armament	\\
55	&	    Hochschule Ausbildung Forschung	&	    Hochschule Studierende Studium	&	universities and \\ & & & higher education	\\
56	&	    Pflege Versorgung Gesundheitswesen	&	    Gesundheitswesen Patientin Patient Patient	&	health care	\\
57	&	    ökologisch Produkt Schutz	&	    Stoff Abfall Chemikalie	&	ecology: waste	\\
58	&	    Landwirtschaft Tier Lebensmittel	&	    Tier Tierschutz Lebensmittel	&	animal protection	\\
59	&	    Mobilität verkehren Straße	&	    Mobilität bahnen ÖPNV	&	mobility	\\
60	&	    Energie Energiewende Verbraucher	&	    Energiewende EEG Strom	&	energy	\\ \hline
\end{tabular}
\caption{Last 30 topics of topic model A trained on German manifestos from 1949 to 2017.}
\label{tab:topic_model_A2}
\end{table}

\begin{table}[t]
\vglue-5mm
\small
	\centering
	\begin{tabular}{|l|c|c|c|} \hline
		number & $\lambda = 1$ & $\lambda = 0.3$ & English label \\ \hline 
1	&	    Gesellschaft Freiheit Politik	&	    Freiheit Gerechtigkeit Gesellschaft	&	freedom 	\\
2	&	    Deutschland Europa Land	&	    sozial Marktwirtschaft Europa Wohlstand	&	social market economy	\\
3	&	    öffentlich Arbeit Investition	&	    Geld Million Investition	&	public investment	\\
4	&	    Deutschland Arbeitsplätze Wirtschaft	&	    Arbeitsplätze Mittelstand Industrie	&	jobs	\\
5	&	    FDP Staat bürgern	&	    FDP bürgern FDP fordern	&	small government	\\
6	&	    Region Land Stadt	&	    ländlich Raum ländlich räumen Region	&	rural areas	\\
7	&	    Deutschland CDU CSU Sport	&	    CDU CSU Sport Ehrenamt	&	sports	\\
8	&	    Arbeit alt Leben	&	    Arbeit Arbeitszeiten Arbeitswelt	&	working hours	\\
9	&	    Kommune Bund Land Ort	&	    Kommune Stadt Gemeinde kommunal	&	local affairs	\\
10	&	    Behinderung Teilhabe Leben	&	    Behinderung Inklusion Barrierefreiheit	&	disability and inclusion	\\
11	&	    Forschung fördern Wirtschaft	&	    Forschung Forschungseinrichtungen Wissenschaft	&	science	\\
12	&	    international BÜNDNIS GRÜNEN setzen	&	    Entwicklungsländer Entwicklungsländern WTO	&	developmental policy	\\
13	&	    setzen Deutschland	&	    Tourismus touristischen	&	tourism	\\
14	&	    Politik sozial PDS	&	    PDS Kapitalismus LINKE	&	anti-capitalism	\\
15	&	    frei Demokrat Leistung Deutschland	&	    frei Demokrat Sozialhilfe Bürgergeld	&	asic income	\\
16	&	    Ostdeutschland Förderung Land	&	    Ostdeutschland ostdeutsch Aufbau Ost	&	east Germany	\\
17	&	    Europa europäisch EU	&	    EU europäisch Mitgliedstaaten	&	refugees and EU	\\
18	&	    Pflege Leistung Pflegeversicherung	&	    Pflege Pflegeversicherung Pflegende	&	care	\\
19	&	    fordern Piratenpartei Piratenpartei setzen	&	    Piratenpartei Piratenpartei setzen Abgeordnete	&	pirates AfD	\\
20	&	    international Bundeswehr Vereinte Nation	&	    Vereinte Nation Entwicklungszusammenarbeit	&	army	\\
    &                                                   & Bundeswehr  & \\
21	&	    Europa Deutschland europäisch	&	    NATO Russland Sicherheitspolitik	&	NATO and Russia	\\
22	&	    Kultur kulturell fördern	&	    Kultur Kunst Kultur kulturell	&	culture	\\
23	&	    Bildung Kind Schule	&	    Schule schulen Schülerin Schüler	&	schools	\\
24	&	    Unternehmen Banken Mitbestimmung	&	    Banken Unternehmen Mitbestimmung	&	finance and banking regulation	\\
25	&	    Arbeit Beschäftigte schaffen	&	    Leiharbeit Mindestlohn Lohndumping	&	work and wages	\\
26	&	    Internet digital Datum	&	    digital Internet Urheberrecht	&	digitalisation	\\
27	&	    Integration Deutschland Gesellschaft	&	    religiös Einbürgerung Religion	&	immigration and religion	\\
28	&	    Hochschule Jugendliche Ausbildung	&	    Hochschule Studierende Studium	&	universities	\\
29	&	    Schutz schützen Sicherheit	&	    Datenschutz Polizei Vorratsdatenspeicherung	&	security and civil rights	\\
30	&	    setzen lehnen fordern	&	    Droge Drogenpolitik Cannabis	&	drugs	\\
31	&	    Rente alt Beitrag	&	    Rente Alterssicherung Rentenversicherung	&	pensions	\\
32	&	    Kind Familie Eltern	&	    Kind Familie Eltern	&	family and children	\\
33	&	    Euro entlasten Steuer	&	    Steuer Steuerreform Einkommensteuer	&	taxes	\\
34	&	    ökologisch Klimaschutz setzen	&	    Naturschutz Natur Wald	&	climate protection	\\
35	&	    Frau Mann Frau Mann	&	    Frau Mann Frau Mann	&	women	\\
36	&	    Versorgung Gesundheitswesen 	&	    Gesundheitswesen Patientin Patient	&	health care	\\
    &       Patientin Patient               &                                           & \\
37	&	    Deutschland lehnen kriegen	&	    Rüstungsexporte Verfassungsschutz Abschiebung	&	arms export and foreign missions	\\
38	&	    Energie Energiewende Mobilität	&	    Energiewende ÖPNV bahnen	&	energy and mobility	\\
39	&	    Wohnung sozial Wohnraum	&	    Wohnung Wohnraum Mieterinnen Mieter	&	housing	\\
40	&	    Landwirtschaft  Verbraucherinnen 	&	Verbraucherinnen Verbraucher Lebensmittel 	&	food safety and animal protection	\\
    &       Verbraucher Lebensmittel   &            Landwirtschaft   & \\ \hline
\end{tabular}
\caption{Topics of topic model B trained on German manifestos from 1989 to 2017. }
\label{tab:topic_model_B}
\end{table}

\end{landscape}
\end{appendix}

\end{document}